\documentclass[journal,twocolumn]{IEEEtran}
\usepackage{}
\usepackage{graphicx}
\usepackage{epstopdf}
\usepackage{amssymb}
\usepackage{amsfonts}
\usepackage{amsmath}
\usepackage{algorithm}
\usepackage{algorithmic}
\usepackage{subeqnarray}
\usepackage{cases}
\usepackage{bm}
\usepackage{color}
\usepackage{subfigure,amsmath,amssymb,cite}
\usepackage{float}
\ifCLASSINFOpdf
\else
\fi
\begin{document}

\title{ Joint Power Allocation and Beamforming for Active IRS-aided Directional Modulation Network}
\author{Rongen Dong, Feng Shu, Yongzhao Li, Jun Li, Yongpeng Wu, and Jiangzhou Wang,\emph{ Fellow, IEEE}

\thanks{Rongen Dong and Feng Shu are with the School of Information and Communication Engineering, Hainan University, Haikou, 570228, China (Email: shufeng0101@163.com).}
\thanks{Yongzhao Li is with School of Telecommunications Engineering, Xidian University, Xi'an, 710071, China (Email:  yzhli@xidian.edu.cn).}
\thanks{Jun Li is with the School of Electronic and Optical Engineering, Nanjing University of Science and Technology, Nanjing, 210094, China (Email: jun.li@njust.edu.cn).}
\thanks{Yongpeng Wu is with the Shanghai Key Laboratory of Navigation and Location Based Services, Shanghai Jiao Tong University, Minhang, Shanghai, 200240, China (Email: yongpeng.wu2016@gmail.com).}
\thanks{Jiangzhou Wang is with the School of Engineering, University of Kent, Canterbury CT2 7NT, U.K. (Email: {j.z.wang}@kent.ac.uk).}

%
}

\maketitle

\begin{abstract}
To boost the secrecy rate (SR) of the conventional directional modulation (DM) network and overcome the double fading effect of the cascaded channels of passive intelligent reflecting surface (IRS), a novel active IRS-assisted DM system with a power adjusting strategy between transmitter and active IRS is proposed in this paper. Then, a joint optimization of maximizing the SR is cast by alternately optimizing the power allocation (PA) factors, transmit beamforming, receive beamforming, and reflect beamforming at IRS, subject to the power constraint at IRS. To tackle the formulated non-convex optimization problem, a high-performance scheme of maximizing SR based on successive convex approximation (SCA) and Schur complement (Max-SR-SS) is proposed, where the derivative operation are employed to optimize the PA factors, the generalized Rayleigh-Rize theorem is adopted to derive the receive beamforming, and the SCA strategy is utilized to design the transmit beamforming and phase shift matrix of IRS. To reduce the high complexity, a low-complexity scheme, named maximizing SR based on equal amplitude reflecting (EAR) and majorization-minimization (MM) (Max-SR-EM), is developed, where the EAR and MM methods are adopted to derive the amplitude and phase of the IRS phase shift matrix, respectively. In particular, when the receivers are single antenna, a scheme of maximizing SR based on alternating optimization (Max-SR-AO) is proposed, where the PA factors, transmit and reflect beamforming are derived by the fractional programming (FP) and SCA algorithms. Simulation results show that with the same power constraint, the SR gains achieved by the proposed schemes outperform those of the fixed PA and passive IRS schemes.

\end{abstract}
\section{Introduction}

The broadcast nature of wireless communication makes the confidential message vulnerable to eavesdropping by the illegal users, leading to security issues of confidential message leakage. Directional modulation (DM), as an advanced and promising physical layer security technology, has attracted the research interest of a wide range of researchers\cite{Daly2010Demonstration, Wan2018Power, shiDOA2022scis,Zhang2020Impact, Dong2022Performance1}.
DM provides security via directive and is suitable for the line-of-sight (LoS) channels such as millimeter wave, unmanned aerial vehicle, intelligent transportation, maritime communication, and satellite communication\cite{Shu2020Directional, Li2020Enabling}. The main ideas of DM are as follows: in the LoS channel, DM transmits confidential message to legitimate user along the desired direction via beamforming vector, and interferes with illegal user eavesdropping by sending artificial noise (AN) in the undesired direction, hence enhancing the secure performance of the system\cite{Cheng2021Physical}. So far, the research for DM technology is mainly focused on the radio frequency frontend and baseband.

To enhance the secrecy rate (SR) of the DM network with a eavesdropper, in \cite{Hong2018Synthesis}, in accordance with the convex optimization method, a sparse array of DM was synthesized, and the proposed approach achieved better flexibility in terms of control security performance and power efficiency. A DM network with hybrid active and passive eavesdroppers was considered in \cite{Qiu2021Security}, and a scheme, which used frequency division array with assisted AN technique at the transmitter to achieve secure transmission with angle-range dependence, was proposed.
Unlike the single legitimate user networks above, the authors in \cite{Kalantari2016Directional} investigated a multi-legitimate user DM network and designed a security-enhancing symbol-level precoding vector, which outperformed the benchmark method in terms of both the power efficiency and security enhancement.
The multi-beam DM networks were investigated in \cite{Hafez2015On} and \cite{Xie2018Artificial}, and a generalized synthesis method and an AN-aided zero-forcing synthesis method were proposed by the former and the latter to enhance the system performance, respectively. However, the above mentioned works mainly focus on the scenario where the legitimate user and the eavesdropper have different directions. To ensure secure transmission of the system when the eavesdropper was in the same direction as the legitimate user, the secure precise wireless transmission DM systems were investigated in \cite{Wu2017Secure} and \cite{Shen2019Two}, which sent confidential message to a specific direction and distance to ensure the secure wireless transmission.

With the development of wireless communication, the demand for network increases dramatically\cite{Pan2020Multicell}. 
Using a large number of active devices will lead to serious energy consumption problems, fortunately, the emergence of intelligent reflecting surface (IRS) provides a novel paradigm to overcome this problem. IRS is a planar array of large numbers of passive electromagnetic elements, each of which is capable of independently adjust the amplitude and phase of the incident signal\cite{Wu2019Intelligent, Yu2020Robust, Liu2022Reconfigurable}. Thanks to this ability, the signal strength at the receiver can be significantly enhanced by properly tuning the reflected signal. Recently, various wireless communication scenarios assisted by IRS have been extensively investigated, including the multicell communications \cite{Pan2020Multicell}, unmanned aerial vehicles communications\cite{Pang2022IRS}, simultaneous wireless information and power transfer (SWIPT) network\cite{Wu2020Joint}, non-orthogonal multiple access network\cite{Fang2020Energy}, and wireless-powered communication network\cite{Hua2022Joint}.

Given the advantages of IRS in wireless communication, in recent years, the IRS-assisted DM network has also been investigated. With the help of IRS, the DM can overcome the limitation of being able to transmit only one confidential bit stream and significantly enhance the SR performance. In \cite{ShuEnhanced2021}, an IRS-aided DM system was considered, and two confidential bit streams were transmitted from Alice to Bob at the same time. Based on the system model of \cite{ShuEnhanced2021}, in \cite{Dong2022Low}, to enhance the SR performance, two low-complexity algorithms were proposed to jointly design the transmit and reflect beamforming vectors of the IRS-assisted DM network.
An IRS-aided DM network equipped with single antenna for both legitimate user and eavesdropper was investigated in \cite{Hu2020Directional}, and the SR closed-form expression was derived. Moreover, the authors in \cite{Lin2023Enhanced} proposed two beamforming algorithms to enhance the SR in the DM network aid by IRS, and they achieved about 30 percent SR gains over no IRS and random phase shift IRS schemes.
The above works showed that the passive IRS can boost the SR performance of the conventional DM network.

However, the ``double fading'' effect that accompanies passive IRS is inevitable, which is caused by the fact that the signal reflected through the IRS needs to pass through the transmitter-to-IRS and IRS-to-receiver cascade links\cite{Zhang2022Active, Liu2022Active, Lin2023WCL}. To overcome this physical limitation, an emerging IRS structure, named active IRS, has been proposed. Unlike the passive IRS, which can only adjust the phase of the incident signal, active IRS integrates active reflection-type amplifiers that can simultaneously tune the amplitude and phase of incident signals. Hence the ``double fading'' effect of the cascaded link can be effectively attenuated, enabling better performance than passive IRS\cite{Zhang2022Active}. Notice that although the active IRS can both amplify and reflect incident signals, it is fundamentally different from full-duplex amplify-and-forward relay. Active IRS does not require radio frequency (RF) chains, has no signal processing capability, and has lower hardware cost\cite{Dong2023Robust}. Moreover, the relay takes two time slots to accomplish the transmission of one signal, whereas active IRS only requires one time slot.

Similar to passive IRS, in recent years, researchers have investigated various wireless communication scenarios with the help of active IRS\cite{Zhu2022Joint}. For example, to maximize the rate of IRS-aided downlink/uplink communication system, the placement of the active IRS was investigated in \cite{You2021Wireless}, which revealed that the system rate was optimal when the active IRS was placed close to the receiver. An active IRS-assisted single input multiple output network was considered in \cite{Long2021Active}, and an alternating optimization approach was proposed to obtain the IRS reflecting coefficient matrix and received beamforming, which achieved the better performance compared to the passive IRS-assisted network with the same power budget. An active IRS-aided SWIPT network was proposed in \cite{Ren2023Transmission}, an alternating iteration method was employed to maximize the weighted sum rate, and the high-performance gain was achieved. The above works presented the benefits of the active IRS for wireless network performance gains.

\begin{figure*}[htbp]
\centering
\includegraphics[width=0.8\textwidth]{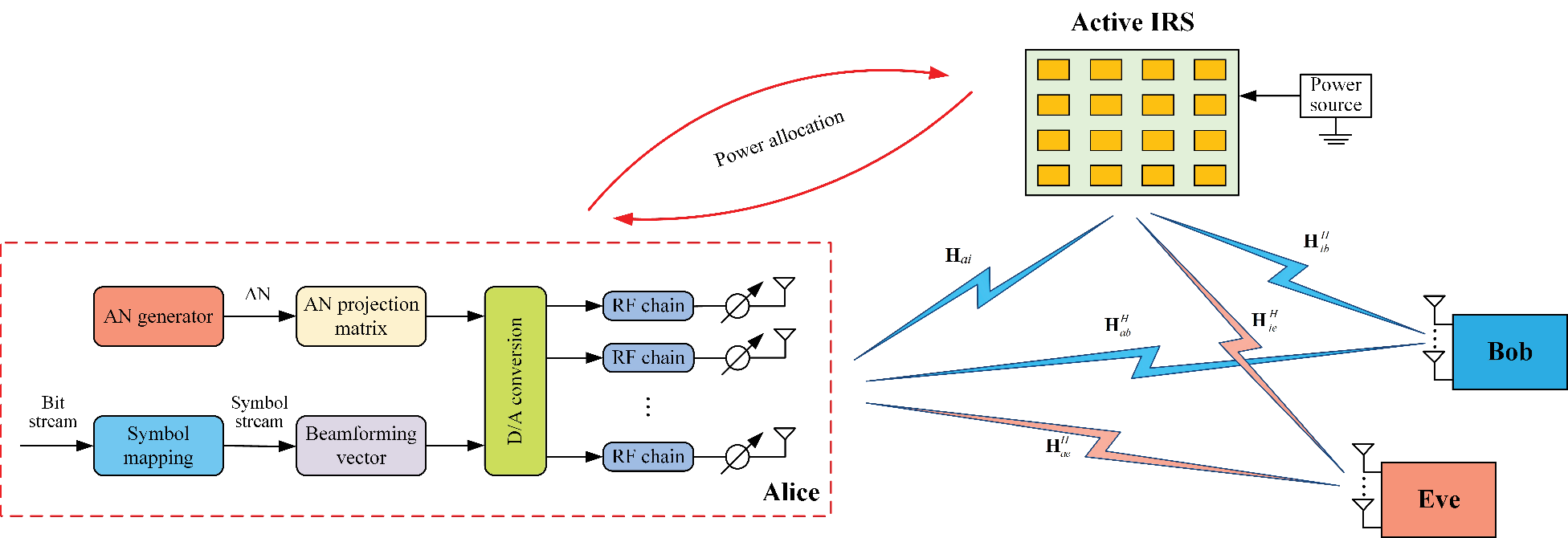}\\
\caption{System diagram of active IRS-assisted DM network.}\label{model}
\end{figure*}

Motivated by the discussions above, to further enhance the SR performance of the passive IRS-assisted DM system, an active IRS-assisted DM network with an eavesdropper is considered in this paper. Given that the beamforming and AN powers of the base station (BS) and IRS power are subject to the system's total power constraint, to investigate the impact of the power allocation (PA) among them and beamforming optimization on the system performance, we focus on maximizing the SR by jointly deriving the PA factors, transmit beamforming, receive beamforming, and reflect beamforming at the active IRS.
To the best of the authors' knowledge, this is the first work to investigate PA between BS and IRS in the active IRS-assisted  secure wireless network. The main contributions of this paper are summarized as follows.
\begin{enumerate}
\item  To enhance the SR performance of the conventional DM system, a novel DM network with the introduction of active IRS is proposed in this paper. Particularly, a PA strategy is proposed to adjust the power fraction between BS and active IRS to further harvest the rate performance gain achieved by active IRS, which does not exist at a passive IRS-aided network. Then, an active IRS-aided DM system with PA is presented. Finally, we formulate a SR maximization problem by jointly optimizing the PA factors, transmit beamforming, receive beamforming, and the IRS phase shift matrix for the active IRS-aided secure DM system in the presence of an eavesdropper, subject to the power constraint at IRS. By optimizing the PA between BS and IRS as well as beamforming, the SR of the system is significantly boosted. 
\item To tackle the formulated non-convex maximum SR optimization problem in which the five variables are coupled with each other, a high-performance alternating optimization scheme, called maximizing SR based on successive convex approximation (SCA) and Schur complement (Max-SR-SS), is proposed. In this scheme, the derivative operation is employed to calculate the optimal PA factor of the confidential message and the PA factor of power allocated to the BS, and the transmit and receive beamforming are derived by the SCA method and the generalized Rayleigh-Rize theorem, respectively, and the phase shift matrix of IRS is calculate by the SCA and Schur complement methods. Moreover, a low-complexity with scheme, named maximizing SR based on equal amplitude reflecting (EAR) and majorization-minimization (MM) (Max-SR-EM), is proposed to address the formulated problem, where the EAR and MM strategies are
    adopted to obtain the amplitude and phase of the IRS phase shift matrix, respectively.
\item In particular, when the receivers are equipped with single antenna, the optimization problem can be simplified and there is no receive beamforming. To tackle the problem, a scheme of maximizing SR based on alternating optimization (Max-SR-AO) is proposed, where the PA factors, transmit beamforming, and phase shift matrix of IRS are designed by the fractional programming (FP) and SCA algorithms. From the simulation results, it is clear that with the same power, the SRs harvested by the proposed three schemes are higher than those of the benchmark schemes. In addition, when the number of phase shift elements tends to large-scale, the gap in terms of SR between the Max-SR-SS and Max-SR-EM schemes is trivial.
\end{enumerate}

The remainder of this paper is organized as follows. We describe the system model of active IRS-assisted DM network and formulate the maximum SR problem in Section \ref{s1}.
Section \ref{s2} introduces the proposed Max-SR-SS and Max-SR-EM schemes.
The proposed Max-SR-AO scheme is described in Section \ref{s3}. The numerical simulation results and conclusions are provided in Section \ref{s4} and Section \ref{s5}, respectively.

{\bf Notations:} in this work, the scalars, vectors and matrices are marked in lowercase, boldface lowercase,  and uppercase letters, respectively. Symbols $(\cdot)^T$, $(\cdot)^*$, $(\cdot)^H$, $\partial(\cdot)$, Tr$(\cdot)$, $(\cdot)^\dag$, $\lambda_{\text{max}}(\cdot)$, $\Re\{\cdot\}$, $\text{diag}\{\cdot\}$, and $\text{blkdiag}\{\cdot\}$ refer to the transpose, conjugate, conjugate transpose, partial derivative, trace, pseudo-inverse, maximum eigenvalue, real part, diagonal, and block diagonal matrix operations, respectively. The sign $|\cdot|$ stands for the scalar's absolute value or the matrix's determinant. The notations $\textbf{I}_Q$ and $\mathbb{C}^{P\times Q}$ mean the identity matrix of $Q\times Q$ and complex-valued matrix space of $P\times Q$, respectively.

\section{system model}\label{s1}

As illustrated in Fig.~1, we investigate an active IRS-assisted secure DM network, where the BS (Alice) sends confidential message to the legitimate user (Bob) with the assistance of active IRS, while sending AN to the eavesdropper (Eve) to reduce the risk of confidential information being intercepted by Eve. There are $N$, $N_b$, and $N_e$ antennas at Alice, Bob, and Eve, respectively. There are $M$ reflection elements on the active IRS with tunable amplitude and phase. In this paper, it is assumed that the active IRS reflects signal only once and there exists the line-of-sight channels. Moreover, all channel state information is assumed to be available owing to the channel estimation.

The transmit signal at Alice is expressed as
\begin{align}
\textbf{s}=\sqrt{\beta l P}\textbf{v}x+\sqrt{(1-\beta) l P}\textbf{T}_{AN}\textbf{z},
\end{align}
where $P$ stands for the total power, $\beta\in(0, 1]$ and $(1-\beta)$ refer to the PA parameters of the confidential message and AN, $l\in (0, 1)$ means the PA factor of the total power allocated to the BS, $\textbf{v}\in \mathbb{C}^{N\times 1}$ and $x$ refer to the beamforming vector and confidential message intent to Bob, they satisfy $\textbf{v}^H\textbf{v}=1$ and $\mathbb{E}[|x|^2]=1$, respectively, $\textbf{T}_{AN}\in \mathbb{C}^{N\times N}$ and $\textbf{z}\in\mathbb{C}^{N\times 1}$ represent the projection matrix and vector of AN, they meet $\text{Tr}(\textbf{T}_{AN}\textbf{T}^H_{AN})=1$ and $\textbf{z}\sim \mathcal {C}\mathcal {N} (\textbf{0}, \textbf{I}_N)$, respectively.

Given the existence of path loss, the received signal at Bob is formulated as
\begin{align}\label{y_b1}
y_b&=\textbf{u}^H_b(\sqrt{g_{ab}}\textbf{H}^H_{ab}+\sqrt{g_{aib}}\textbf{H}^H_{ib}\boldsymbol{\Psi}\textbf{H}_{ai})\textbf{s}+
\sqrt{g_{ib}}\textbf{u}^H_b\textbf{H}^H_{ib}\boldsymbol{\Psi}\textbf{n}_r\nonumber\\
&~~~+n_b\nonumber\\
&=\sqrt{\beta l P}\textbf{u}^H_b
(\sqrt{g_{ab}}\textbf{H}^H_{ab}+\sqrt{g_{aib}}\textbf{H}^H_{ib}\boldsymbol{\Psi}\textbf{H}_{ai})\textbf{v}x+\nonumber\\
&~~~\sqrt{(1-\beta) l P}\textbf{u}^H_b(\sqrt{g_{ab}}\textbf{H}^H_{ab}+\sqrt{g_{aib}}\textbf{H}^H_{ib}\boldsymbol{\Psi}\textbf{H}_{ai})
\textbf{T}_{AN}\textbf{z}+\nonumber\\
&~~~\sqrt{g_{ib}}\textbf{u}^H_b\textbf{H}^H_{ib}\boldsymbol{\Psi}\textbf{n}_r+n_b,
\end{align}
where $\textbf{u}_b\in \mathbb{C}^{N_b\times 1}$ refers to the receive beamforming, $g_{ab}$ and $g_{ib}$ stand for the path loss parameters of Alice-to-Bob and IRS-to-Bob channels, respectively, $g_{aib}=g_{ai}g_{ib}$ means the equivalent path loss parameter of Alice-to-IRS and IRS-to-Bob channels, $\boldsymbol{\Psi}=\text{diag}\{\psi_1, \cdots, \psi_m, \cdots, \psi_M\}\in \mathbb{C}^{M\times M}$ and $\boldsymbol{\psi}=[\psi_1, \cdots, \psi_m, \cdots, \psi_M]^T\in \mathbb{C}^{M\times 1}$ refer to the reflection coefficient matrix and vector of the active IRS, $\psi_m=\alpha_me^{j{\phi}_m}$, $\alpha_m$ and ${\phi}_m$ are the amplitude and phase of $m$-th reflecting element, respectively. $\textbf{n}_r \sim \mathcal {C}\mathcal {N} (\textbf{0}, \sigma^2_r\textbf{I}_M)$ and $n_b \sim \mathcal {C}\mathcal {N} (0, \sigma^2_b)$ mean the complex additive white Gaussian noise (AWGN) at IRS and at Bob, respectively, $\textbf{H}^H_{ab}=\textbf{h}_{ba}\textbf{h}^H_{ab}\in \mathbb{C}^{N_b\times N}$, $\textbf{H}^H_{ib}=\textbf{h}_{bi}\textbf{h}^H_{ib}\in \mathbb{C}^{N_b\times M}$, and $\textbf{H}_{ai}=\textbf{h}_{ia}\textbf{h}^H_{ai}\in \mathbb{C}^{M\times N}$ denote the Alice-to-Bob, IRS-to-Bob, and Alice-to-IRS channels, respectively. It is assumed that $\textbf{h}_{tr}=\textbf{h}(\theta_{tr})$ for simplicity, and the normalized steering vector is
\begin{align}\label{h_theta}
\textbf{h}(\theta)\buildrel \Delta \over=\frac{1}{\sqrt{N}}[e^{j2\pi\Phi_{\theta}(1)}, \dots, e^{j2\pi\Phi_{\theta}(n)}, \dots, e^{j2\pi\Phi_{\theta}(N)}]^T,
\end{align}
where
\begin{align}
\Phi_{\theta}(n)=-\Big(n-\frac{N+1}{2}\Big)\frac{d \cos\theta}{\lambda}, n=1, 2, \dots, N,
\end{align}
$\theta$ represents the direction angle of the signal departure or arrival, $n$ stands for the antenna index, $d$ indicates the distance between adjacent transmitting antennas, and $\lambda$ refers to the wavelength.

Similarly, the received signal at Eve is cast as
\begin{align}\label{y_e1}
y_e&=\textbf{u}^H_e(\sqrt{g_{ae}}\textbf{H}^H_{ae}+\sqrt{g_{aie}}\textbf{H}^H_{ie}\boldsymbol{\Psi}\textbf{H}_{ai})\textbf{s}+
\sqrt{g_{ie}}\textbf{u}^H_e\textbf{H}^H_{ie}\boldsymbol{\Psi}\textbf{n}_r\nonumber\\
&~~~+n_e\nonumber\\
&=\sqrt{\beta l P}\textbf{u}^H_e
(\sqrt{g_{ae}}\textbf{H}^H_{ae}+\sqrt{g_{aie}}\textbf{H}^H_{ie}\boldsymbol{\Psi}\textbf{H}_{ai})\textbf{v}x+\nonumber\\
&~~~\sqrt{(1-\beta) l P}\textbf{u}^H_e(\sqrt{g_{ae}}\textbf{H}^H_{ae}+\sqrt{g_{aie}}\textbf{H}^H_{ie}\boldsymbol{\Psi}\textbf{H}_{ai})
\textbf{T}_{AN}\textbf{z}+\nonumber\\
&~~~\sqrt{g_{ie}}\textbf{u}^H_e\textbf{H}^H_{ie}\boldsymbol{\Psi}\textbf{n}_r+n_e,
\end{align}
where $\textbf{u}_e\in \mathbb{C}^{N_e\times 1}$ denotes the receive beamforming, $g_{ae}$ and $g_{ie}$ stand for the path loss parameters of Alice-to-Eve and IRS-to-Eve channels, respectively, $g_{aie}=g_{ai}g_{ie}$ means the equivalent path loss parameter of Alice-to-IRS and IRS-to-Eve channels, $n_e$ represents the AWGN at Eve that satisfies the distribution $n_e \sim \mathcal {C}\mathcal {N} (0, \sigma^2_e)$, $\textbf{H}^H_{ae}=\textbf{h}_{ea}\textbf{h}^H_{ae}\in \mathbb{C}^{N_e\times N}$ and $\textbf{H}^H_{ie}=\textbf{h}_{ei}\textbf{h}^H_{ie}\in \mathbb{C}^{1\times M}$ refer to the Alice-to-Eve and IRS-to-Eve channels, respectively.

It is assumed that AN is transmitted to Eve for jamming eavesdropping only and does not impact Bob, based on the criterion of null-space projection,  $\textbf{T}_{AN}$ should meet
\begin{align}\label{}
\textbf{H}_{ai}\textbf{T}_{AN}=\textbf{0}_{M\times N},~ \textbf{H}^H_{ab}\textbf{T}_{AN}=\textbf{0}_{N_b\times N}.
\end{align}
Let us define a equivalent virtual channel matrix of confidential message as follows
\begin{align}
\textbf{H}_{CM}=\left[ \begin{array}{*{20}{c}}
\textbf{H}_{ai}\\
\textbf{H}^H_{ab}
\end{array}\right]_{(M+N_b)\times N}.
\end{align}
Then, $\textbf{T}_{AN}$ can be designed as
\begin{align}
\textbf{T}_{AN}=\textbf{I}_N-\textbf{H}_{CM}^H[\textbf{H}_{CM}\textbf{H}_{CM}^H]^\dag \textbf{H}_{CM}.
\end{align}
At this point, (\ref{y_b1}) and (\ref{y_e1}) can be rewritten as
\begin{align}\label{y_b2}
y_b&=\sqrt{\beta l P}\textbf{u}^H_b
\left(\sqrt{g_{ab}}\textbf{H}^H_{ab}+\sqrt{g_{aib}}\textbf{H}^H_{ib}\boldsymbol{\Psi}\textbf{H}_{ai}\right)\textbf{v}x+\nonumber\\
&~~~\sqrt{g_{ib}}\textbf{u}^H_b\textbf{H}^H_{ib}\boldsymbol{\Psi}\textbf{n}_r+n_b,
\end{align}
and
\begin{align}\label{y_e2}
y_e&=\sqrt{\beta l P}\textbf{u}^H_e\left(\sqrt{g_{ae}}\textbf{H}^H_{ae}+
\sqrt{g_{aie}}\textbf{H}^H_{ie}\boldsymbol{\Psi}\textbf{H}_{ai}\right)\textbf{v}x+\nonumber\\
&~~~\sqrt{(1-\beta) l P}\sqrt{g_{ae}}\textbf{u}^H_e\textbf{H}^H_{ae}\textbf{T}_{AN}\textbf{z}+
\sqrt{g_{ie}}\textbf{u}^H_e\textbf{H}^H_{ie}\boldsymbol{\Psi}\textbf{n}_r\nonumber\\
&~~~+n_e,
\end{align}
respectively.

Based on (\ref{y_b2}) and (\ref{y_e2}), the achievable rates at Bob and Eve are given by
\begin{align}\label{y_b3}
R_b=\text{log}_2\left(1+\frac{\beta l P|\textbf{u}^H_b\left(\sqrt{g_{ab}}\textbf{H}^H_{ab}+
\sqrt{g_{aib}}\textbf{H}^H_{ib}\boldsymbol{\Psi}\textbf{H}_{ai}\right)\textbf{v}|^2}
{\sigma^2_r\|\sqrt{g_{ib}}\textbf{u}^H_b\textbf{H}^H_{ib}\boldsymbol{\Psi}\|^2+\sigma^2_b}\right),
\end{align}
and
\begin{align}\label{y_e31}
&R_e=\text{log}_2\left(1+\gamma\right),
\end{align}
respectively, where
\begin{align}
&\gamma=\nonumber\\
&\frac{\beta l P|\textbf{u}^H_e\left(\sqrt{g_{ae}}\textbf{H}^H_{ae}+
\sqrt{g_{aie}}\textbf{H}^H_{ie}\boldsymbol{\Psi}\textbf{H}_{ai}\right)\textbf{v}|^2}
{(1-\beta) l P g_{ae}\|\textbf{u}^H_e\textbf{H}^H_{ae}\textbf{T}_{AN}\|^2+
\sigma^2_r\|\sqrt{g_{ie}}\textbf{u}^H_e\textbf{H}^H_{ie}\boldsymbol{\Psi}\|^2+\sigma^2_e}.
\end{align}
Due to the fact that Alice and Bob cannot capture Eve's received beamforming $\textbf{u}_e$ in general, a upper bound of (\ref{y_e31}) can be obtained by
\begin{align}\label{y_e3}
R_e&\leq\text{log}_2(1+\text{Tr}((1-\beta) l Pg_{ae}\textbf{H}^H_{ae}\textbf{T}_{AN}\textbf{T}^H_{AN}\textbf{H}_{ae}+\nonumber\\
&~~~\sigma^2_rg_{ie}\textbf{H}^H_{ie}\boldsymbol{\Psi}\boldsymbol{\Psi}^H\textbf{H}_{ie}+\sigma^2_e\textbf{I}_{N_e})^{-1}(\beta l P(\sqrt{g_{ae}}\textbf{H}^H_{ae}+\nonumber\\
&~~~\sqrt{g_{aie}}\textbf{H}^H_{ie}\boldsymbol{\Psi}\textbf{H}_{ai})\textbf{v}\textbf{v}^H(\sqrt{g_{ae}}\textbf{H}^H_{ae}+\sqrt{g_{aie}}\textbf{H}^H_{ie}\boldsymbol{\Psi}\textbf{H}_{ai})^H))\nonumber\\
&\buildrel \Delta \over=\widetilde{R}_e.
\end{align}
The detailed derivation is available in Appendix.

At this point, the lower bound of SR for the system is expressed as
\begin{align}
R_s=\text{max}\{0, R_b-\widetilde{R}_e\}.
\end{align}

Moreover, the transmitted power at active IRS can be formulated as follows
\begin{align}
P_r=\text{Tr}\left(\boldsymbol{\Psi}(g_{ai}\beta l P\textbf{H}_{ai}\textbf{v}\textbf{v}^H\textbf{H}_{ai}^H+\sigma^2_r\textbf{I}_M)\boldsymbol{\Psi}^H\right).
\end{align}

In this paper, we maximize the SR by jointly deriving the PA factors $\beta$ and $l$, transmit beamforming $\textbf{v}$, receive beamforming $\textbf{u}_b$, and active IRS phase shift matrix $\boldsymbol{\Psi}$. The overall optimization problem is formulated as follows
\begin{subequations}\label{p0}
\begin{align}
&\max \limits_{\beta, l, \textbf{v}, \textbf{u}_b, \boldsymbol{\Psi}}
~~R_s\\
&~~~~\text{s.t.} ~~~~\textbf{v}^H\textbf{v}=1, ~\textbf{u}_b^H\textbf{u}_b=1,\\
& ~~~~~~~~~~~0<\beta\leq 1,~0<l< 1,\\
& ~~~~~~~~~~~|\boldsymbol{\Psi}(m,m)|\leq {\psi}^{\text{max}}, m=1,\dots, M,\\
& ~~~~~~~~~~~P_r\leq (1-l)P,\label{P_r}
\end{align}
\end{subequations}
where ${\psi}^{\text{max}}$ means the amplification gain threshold of the active IRS elements, and $(1-l)P$ refers to the maximum transmit power of the active IRS. It is obvious that this optimization problem has a non-convex objective function and constraints, and the optimization variables are highly coupled with each other, which makes it a challenge to address it directly in general. Hence, the alternating iteration strategy is taken into account for solving this optimization problem in what follows.

\section{Proposed Max-SR-SS and Max-SR-EM schemes}\label{s2}
In this section, to streamline the solution of the problem, we aim at maximizing SR and decompose the problem (\ref{p0}) into five subproblems. In what follows, the parameters $\beta$, $l$, $\textbf{v}$, $\textbf{u}_b$, and $\boldsymbol{\Psi}$ are sequentially optimized by fixing the other variables.

\subsection{Optimization of the PA factor $\beta$}
In this subsection, the transmit beamforming $\textbf{v}$, receive beamforming $\textbf{u}_b$, and IRS phase shift matrix $\boldsymbol{\Psi}$ are given for the sake of simplicity, we re-arrange the IRS power constraint (\ref{P_r}) as
\begin{align}\label{beta1}
\beta l \text{Tr}\left(\boldsymbol{\Psi}(g_{ai} P\textbf{H}_{ai}\textbf{v}\textbf{v}^H\textbf{H}_{ai}^H)\boldsymbol{\Psi}^H\right)+
\text{Tr}(\sigma^2_r\boldsymbol{\Psi}\boldsymbol{\Psi}^H)\leq (1-l)P.
\end{align}

For the sake of simplicity, let us define
\begin{subequations}
\begin{align}
&A_b=P|\textbf{u}^H_b(\sqrt{g_{ab}}\textbf{H}^H_{ab}+
\sqrt{g_{aib}}\textbf{H}^H_{ib}\boldsymbol{\Psi}\textbf{H}_{ai})\textbf{v}|^2,\\
&B_b=\sigma^2_r\|\sqrt{g_{ib}}\textbf{u}^H_b\textbf{H}^H_{ib}\boldsymbol{\Psi}\|^2+\sigma^2_b.
\end{align}
\end{subequations}
Then, (\ref{y_b3}) can be degenerated to
\begin{align}\label{Rb0}
&R_b=\text{log}_2\left(\frac{\beta l A_b+B_b}{B_b}\right).
\end{align}

Let us define
\begin{align}
&\textbf{A}=P(\sqrt{g_{ae}}\textbf{H}^H_{ae}+\sqrt{g_{aie}}\textbf{H}^H_{ie}\boldsymbol{\Psi}\textbf{H}_{ai})\textbf{v}\textbf{v}^H\cdot\nonumber\\
&~~~~~~(\sqrt{g_{ae}}\textbf{H}^H_{ae}+\sqrt{g_{aie}}\textbf{H}^H_{ie}\boldsymbol{\Psi}\textbf{H}_{ai})^H,\\
&\textbf{B}=\sigma^2_rg_{ie}\textbf{H}^H_{ie}\boldsymbol{\Psi}\boldsymbol{\Psi}^H\textbf{H}_{ie}+\sigma^2_e\textbf{I}_{N_e},
\end{align}
and based on
\begin{align}
&(1-\beta)lPg_{ae}\textbf{H}_{ae}^H\textbf{T}_{AN}\textbf{T}_{AN}^H\textbf{H}_{ae}\nonumber\\
&=(1-\beta)l\underbrace{Pg_{ae}\textbf{H}_{ae}^H\textbf{T}_{AN}\textbf{T}_{AN}^H\textbf{h}_{ea}}_{\textbf{q}}\textbf{h}^H_{ae},
\end{align}
(\ref{y_e3}) can be reformulated as
\begin{align}\label{Re22}
&\widetilde{R}_e=\text{log}_2(1+\text{Tr}[(\textbf{B}+(1-\beta)l\textbf{q}\textbf{h}^H_{ae})^{-1}(\beta l \textbf{A})]).
\end{align}
Due to the presence of inverse operation, the Sherman-Morrison theorem is taken into account for the simplification, i.e.,
\begin{align}
(\textbf{Z}+\textbf{x}\textbf{y}^T)^{-1}=\textbf{Z}^{-1}-\frac{\textbf{Z}^{-1}\textbf{x}\textbf{y}^T\textbf{Z}^{-1}}{\textbf{y}^T\textbf{Z}^{-1}\textbf{x}+1},
\end{align}
then, we have
\begin{align}
(\textbf{B}+(1-\beta)l\textbf{q}\textbf{h}^H_{ae})^{-1}=\textbf{B}^{-1}-\frac{(1-\beta)l\textbf{B}^{-1}\textbf{q}\textbf{h}^H_{ae}\textbf{B}^{-1}}
{(1-\beta)l\textbf{h}^H_{ae}\textbf{B}^{-1}\textbf{q}+1},
\end{align}
and (\ref{Re22}) becomes
\begin{align}
&\widetilde{R}_e=\text{log}_2\Big(1+\beta l\text{Tr}[\textbf{B}^{-1}\textbf{A}]-\frac{\beta(1-\beta)l^2\text{Tr}[\textbf{B}^{-1}\textbf{q}\textbf{h}^H_{ae}\textbf{B}^{-1}\textbf{A}]}{(1-\beta)l\textbf{h}^H_{ae}\textbf{B}^{-1}\textbf{q}+1}\Big).
\end{align}
Let us define
$A_e=\text{Tr}[\textbf{B}^{-1}\textbf{A}],$
$B_e=\text{Tr}[\textbf{B}^{-1}\textbf{q}\textbf{h}^H_{ae}\textbf{B}^{-1}\textbf{A}],$
$C_e=\textbf{h}^H_{ae}\textbf{B}^{-1}\textbf{q}$.
Then, (\ref{y_e3}) can be recast as
\begin{align}\label{Re0}
&\widetilde{R}_e=\nonumber\\
&\text{log}_2\left(\frac{(1-\beta)l C_e+1+\beta(1-\beta)l^2(A_eC_e-B_e)+\beta l A_e}{(1-\beta)l C_e+1}\right),
\end{align}
respectively.

In what follows, we handle the optimization of the PA parameters $\beta$ and $l$ successively.

Defining
$E_1=l^2(A_eC_e -B_e),$
$E_2=l^2(A_eC_e-B_e)-lC_e+lA_e,$
$E_3=lC_e+1,$
$E_4=lC_e.$
Given $l$, in accordance with (\ref{p0}), (\ref{Rb0}), and (\ref{Re0}), the optimization problem with respect to $\beta$ can be simplified as follows
\begin{subequations}\label{beta20}
\begin{align}
&\max \limits_{\beta}
~~f_1(\beta)=\frac{\beta^2A_1-\beta B_1-C_1}{\beta^2 D_1-\beta F_1-C_1}\label{object}\\
&~~\text{s.t.} ~~~\beta K_1\leq L_1, 0<\beta\leq 1,
\end{align}
\end{subequations}
where
$A_1=lA_bE_4, $
$B_1=lA_bE_3-B_bE_4,$
$C_1=B_bE_3, $
$D_1=E_1B_b,$
$F_1=E_2B_b, $
$K_1=l \text{Tr}\left(\boldsymbol{\Psi}(g_{ai} P\textbf{H}_{ai}\textbf{v}\textbf{v}^H\textbf{H}_{ai}^H)\boldsymbol{\Psi}^H\right),$
$L_1= (1-l)P-\text{Tr}(\sigma^2_r\boldsymbol{\Psi}\boldsymbol{\Psi}^H)$.
Then, (\ref{beta20}) can be reformulated as
\begin{subequations}\label{beta22}
\begin{align}
&\max \limits_{\beta}
~~f_1(\beta)=\frac{\beta^2A_1-\beta B_1-C_1}{\beta^2 D_1-\beta F_1-C_1}\label{object}\\
&~~\text{s.t.} ~~~0<\beta\leq \beta^{\text{max}},
\end{align}
\end{subequations}
where $\beta^{\text{max}}\buildrel \Delta \over=\text{min}\big\{\frac{L_1}{K_1},1 \big\}$. Given that the denominator $\beta^2 D_1-\beta F_1-C_1\neq0$,
we can obtain that the objective function of problem (\ref{beta22}) is continuous and differentiable in the interval
$(0, \beta^{\text{max}}]$. Then, we take its partial derivative and make it equal to 0 yields
\begin{align}
\frac{\partial f_1(\beta)}{\partial \beta}&=\frac{1}{(\beta^2 D_1-\beta F_1-C_1)^2}\big[\beta^2(B_1D_1-A_1F_1)+\nonumber\\
&~~~2\beta (C_1D_1-A_1C_1)+(B_1C_1-C_1F_1)\big]\nonumber\\
&=0,
\end{align}
which can can be simplified as
\begin{align}\label{beta1}
&\beta^2(B_1D_1-A_1F_1)+2\beta (C_1D_1-A_1C_1)+(B_1C_1-C_1F_1)\nonumber\\
&=0,
\end{align}
\subsubsection{When $B_1D_1-A_1F_1\neq 0$}
the equation (\ref{beta1}) is a quadratic. Let us define
\begin{align}
\Delta_{\beta}=4(C_1D_1-A_1C_1)^2-4(B_1D_1-A_1F_1)(B_1C_1-C_1F_1).
\end{align}
if $\Delta_{\beta}\geq 0$, based on the formula for the roots of a quadratic function, we can get its roots as
\begin{align}
&\beta_1=\frac{-2(C_1D_1-A_1C_1)+\sqrt{\Delta_{\beta}}}{2(B_1D_1-A_1F_1)}, \\
&\beta_2=\frac{-2(C_1D_1-A_1C_1)-\sqrt{\Delta_{\beta}}}{2(B_1D_1-A_1F_1)}.
\end{align}
\subsubsection{When $B_1D_1-A_1F_1= 0$}
(\ref{beta1}) can be degraded to
\begin{align}\label{}
2\beta (C_1D_1-A_1C_1)+(B_1C_1-C_1F_1)=0,
\end{align}
which yields
\begin{align}
\beta_3=-\frac{B_1-F_1}{2(D_1-A_1)}.
\end{align}

Next, we judge whether these candidate solutions of $\beta$ are in the interval $(0, \beta^{\text{max}}]$. Finally, the optimal value of $\beta$ can be obtained by comparing the values of $f_1(\beta)$ at endpoints and candidate solutions.
The detailed procedures for deriving the PA factor $\beta$ is shown in Algorithm 1.
\begin{algorithm}
\caption{The algorithm for optimizing $\beta$}
\begin{algorithmic}[1] 
\STATE If $B_1D_1-A_1F_1\neq 0$ and $\Delta_{\beta}\geq 0$, the four different scenarios are considered as follows.
\begin{enumerate}
\item If $\beta_1, \beta_2 \in (0, \beta^{\text{max}}]$, then compare the values of $f_1(0)$, $f_1(\beta_1)$, $f_1(\beta_2)$, and $f_1(\beta^{\text{max}})$.
\item If $\beta_1\in (0, \beta^{\text{max}}]$ and $\beta_2 \notin (0, \beta^{\text{max}}]$, then compare the values of $f_1(0)$, $f_1(\beta_1)$, and $f_1(\beta^{\text{max}})$.
\item If $\beta_1 \notin (0, \beta^{\text{max}}]$ and $\beta_2 \in (0, \beta^{\text{max}}]$, then compare the values of $f_1(0)$, $f_1(\beta_2)$, and $f_1(\beta^{\text{max}})$.
\item If $\beta_1, \beta_2 \notin (0, \beta^{\text{max}}]$, then compare the values of $f_1(0)$ and $f_1(\beta^{\text{max}})$.
\end{enumerate}
\STATE If $B_1D_1-A_1F_1\neq 0$ and $\Delta_{\beta}< 0$, the optimal PA parameter has been shown in aforementioned 4).
\STATE If $B_1D_1-A_1F_1=0$, the two different scenarios are taken into account as follows.
\begin{enumerate}
\item If $\beta_3 \in (0, \beta^{\text{max}}]$, then compare the values of $f_1(0)$, $f_1(\beta_3)$, and $f_1(\beta^{\text{max}})$.
\item If $\beta_3 \notin (0, \beta^{\text{max}}]$, then compare the values of $f_1(0)$ and $f_1(\beta^{\text{max}})$.
\end{enumerate}
\STATE Output the optimal PA factor $\beta^{\text{opt}}$.
\end{algorithmic}
\end{algorithm}

\subsection{Optimization of the PA factor $l$}
Fixed $\textbf{v}$, $\textbf{u}_b$, and $\boldsymbol{\Psi}$, given that the optimal $\beta$ has been found in the previous subsection, we transfer the focus to solving for $l$. Let us define
$E_5=\beta(1-\beta)(A_eC_e-B_e),$
$E_6=(1-\beta)C_e+\beta A_e,$
$E_7=(1-\beta)C_e.$
In accordance with (\ref{Rb0}) and (\ref{Re0}), by neglecting the constant terms, the optimization problem with respect to $l$ can be simplified as follows
\begin{subequations}\label{L2}
\begin{align}
&\max \limits_{l}
~~f_2(l)=\frac{l^2A_2+lB_2+C_2}{l^2D_2+lF_2+C_2}\\
&~~\text{s.t.} ~~~l K_2\leq L_2, 0<l< 1,
\end{align}
\end{subequations}
where
$A_2=\beta A_bE_7,$
$B_2=\beta A_b+E_7B_b,$
$C_2=B_b,$
$D_2=E_5B_b,$
$F_2=E_6B_b,$
$K_2=\beta \text{Tr}\left(\boldsymbol{\Psi}(g_{ai} P\textbf{H}_{ai}\textbf{v}\textbf{v}^H\textbf{H}_{ai}^H)\boldsymbol{\Psi}^H\right)+P,$
$L_2=P-\text{Tr}(\sigma^2_r\boldsymbol{\Psi}\boldsymbol{\Psi}^H)$.
Further simplification yields
\begin{subequations}\label{L00}
\begin{align}
&\max \limits_{l}
~~f_2(l)=\frac{l^2A_2+lB_2+C_2}{l^2D_2+lF_2+C_2}\\
&~~\text{s.t.} ~~~0<l\leq l^{\text{max}},
\end{align}
\end{subequations}
where $l^{\text{max}}\buildrel \Delta \over=\text{min}\big\{\frac{L_2}{K_2}, 1\big\}$. Due to the fact that the denominator $l^2D_2+lF_2+C_2\neq0$,
we can obtain that the objective function of problem (\ref{L00}) is continuous and differentiable in the interval
$(0, l^{\text{max}}]$. Then, we take its partial derivative and make it equal to 0 yields
\begin{align}
\frac{\partial f_2(l)}{\partial l}&=\frac{1}{(l^2D_2+lF_2+C_2)^2}\big[l^2(A_2F_2-B_2D_2)+\nonumber\\
&~~~2l(A_2C_2-C_2D_2)+(B_2C_2-C_2F_2)\big]\nonumber\\
&=0,
\end{align}
which yields
\begin{align}\label{L0}
&l^2(A_2F_2-B_2D_2)+2l(A_2C_2-C_2D_2)+(B_2C_2-C_2F_2)\nonumber\\
&=0.
\end{align}
\subsubsection{When $A_2F_2-B_2D_2\neq 0$}
the equation (\ref{L0}) is a quadratic. Let us define
\begin{align}
\Delta_{l}=4(A_2C_2-C_2D_2)^2-4(A_2F_2-B_2D_2)(B_2C_2-C_2F_2).
\end{align}
if $\Delta_{l}\geq 0$, based on the formula for the roots of a quadratic function, we can get its roots as
\begin{align}
&l_1=\frac{-2(A_2C_2-C_2D_2)+\sqrt{\Delta_{l}}}{2(A_2C_2-C_2D_2)}, \\
&l_2=\frac{-2(A_2C_2-C_2D_2)-\sqrt{\Delta_{l}}}{2(A_2C_2-C_2D_2)}.
\end{align}
\subsubsection{When $A_2F_2-B_2D_2= 0$}
(\ref{L0}) can be recast as 
\begin{align}\label{}
2l(A_2C_2-C_2D_2)+(B_2C_2-C_2F_2)=0,
\end{align}
we have
\begin{align}
l_3=-\frac{B_2-F_2}{2(A_2-D_2)}.
\end{align}

Next, an analysis similar to solving for $\beta$ needs to be performed, and we ignore the procedure for the sake of avoiding repetition.

\subsection{Optimization of the transmit beamforming vector $\textbf{v}$}
Given $\beta$, $l$, $\textbf{u}_b$, and $\boldsymbol{\Psi}$, we reformulate the IRS power constraint (\ref{P_r}) as follows
\begin{align}\label{P_r1}
P_r=\textbf{v}^H(g_{ai}\beta l P\textbf{H}^H_{ai}\boldsymbol{\Psi}^H\boldsymbol{\Psi}\textbf{H}_{ai})\textbf{v}+
\text{Tr}(\sigma^2_r\boldsymbol{\Psi}\boldsymbol{\Psi}^H)\leq(1-l)P.
\end{align}
With ignoring the constant term, (\ref{p0}) can be re-arranged as the optimization problem with respect to $\textbf{v}$ as follows
\begin{subequations}\label{v1}
\begin{align}
&\max \limits_{\textbf{v}}
~~\frac{\textbf{v}^H\textbf{C}\textbf{v}}{\textbf{v}^H\textbf{D}\textbf{v}}\\
&~~\text{s.t.} ~~~\textbf{v}^H\textbf{v}=1, (\ref{P_r1}),
\end{align}
\end{subequations}
where
\begin{align}
\textbf{C}&=\beta l P\left(\sqrt{g_{ab}}\textbf{H}^H_{ab}+
\sqrt{g_{aib}}\textbf{H}^H_{ib}\boldsymbol{\Psi}\textbf{H}_{ai}\right)^H\textbf{u}_b\textbf{u}^H_b\big(\sqrt{g_{ab}}\textbf{H}^H_{ab}+\nonumber\\
&~~~\sqrt{g_{aib}}\textbf{H}^H_{ib}\boldsymbol{\Psi}\textbf{H}_{ai}\big)/\left(\sigma^2_r\|\sqrt{g_{ib}}\textbf{u}_b^H\textbf{H}^H_{ib}\boldsymbol{\Psi}\|^2+\sigma^2_b\right)
+\textbf{I}_N,\\
\textbf{D}&=\beta l P\left(\sqrt{g_{ae}}\textbf{H}^H_{ae}+\sqrt{g_{aie}}\textbf{H}^H_{ie}\boldsymbol{\Psi}\textbf{H}_{ai}\right)^H\big[(1-\beta) l P g_{ae}\textbf{H}^H_{ae}\cdot\nonumber\\
&~~\textbf{T}_{AN}\textbf{T}^H_{AN}\textbf{H}_{ae}+\sigma^2_rg_{ie}\textbf{H}^H_{ie}\boldsymbol{\Psi}\boldsymbol{\Psi}^H\textbf{H}_{ie}
+\sigma^2_e\textbf{I}_{N_e}\big]^{-1}\big(\sqrt{g_{ae}}\textbf{H}^H_{ae}\nonumber\\
&~~+\sqrt{g_{aie}}\textbf{H}^H_{ie}\boldsymbol{\Psi}\textbf{H}_{ai}\big)
+\textbf{I}_N.
\end{align}

Given that the objective function value in (\ref{v1}) is insensitive to the scaling of $\textbf{v}$, we relax the equation constraint to $\textbf{v}^H\textbf{v}\leq1$\cite{ShuEnhanced2021}. Then, in accordance with the first order Taylor approximation, we have
\begin{align}\label{T}
\frac{|y|^2}{z}\geq -\frac{\bar{y}^*\bar{y}}{\bar{z}^2}z+\frac{2\Re\{\bar{y}^*y\}}{\bar{z}}.
\end{align}
Then, the problem (\ref{v1}) can be recast as
\begin{subequations}\label{v33}
\begin{align}
&\max \limits_{\textbf{v}}
~~-\frac{\bar{\textbf{v}}^H\textbf{C}\bar{\textbf{v}}}{(\bar{\textbf{v}}^H\textbf{D}\bar{\textbf{v}})^2}
\textbf{v}^H\textbf{D}\textbf{v}+\frac{2\Re\{\bar{\textbf{v}}^H\textbf{C}\textbf{v}\}}
{\bar{\textbf{v}}^H\textbf{D}\bar{\textbf{v}}}\\
&~~\text{s.t.} ~~~\textbf{v}^H\textbf{v}\leq1, (\ref{P_r1}),
\end{align}
\end{subequations}
where $\bar{\textbf{v}}$ stands for the given vector.
This is a convex optimization problem that can be tackled directly with convex optimizing toolbox (e.g. CVX\cite{Grant2012CVX}).

\subsection{Optimization of the receive beamforming vector $\textbf{u}_b$}
Fixed $\beta$, $l$, $\textbf{v}$, and $\boldsymbol{\Psi}$, the optimization problem with respect to $\textbf{u}_b$ can be re-arranged as
\begin{subequations}\label{u_b}
\begin{align}
&\max \limits_{\textbf{u}_b}
~~\frac{\textbf{u}^H_b\textbf{A}_1\textbf{u}_b}{\textbf{u}^H_b\textbf{A}_2\textbf{u}_b}\\
&~~\text{s.t.} ~~~\textbf{u}^H_b\textbf{u}_b=1,
\end{align}
\end{subequations}
where
\begin{align}
&\textbf{A}_1=\beta l P\left(\sqrt{g_{ab}}\textbf{H}^H_{ab}+
\sqrt{g_{aib}}\textbf{H}^H_{ib}\boldsymbol{\Psi}\textbf{H}_{ai}\right)\textbf{v}\textbf{v}^H\cdot\nonumber\\
&~~~~~~\left(\sqrt{g_{ab}}\textbf{H}^H_{ab}+\sqrt{g_{aib}}\textbf{H}^H_{ib}\boldsymbol{\Psi}\textbf{H}_{ai}\right)^H,\\
&\textbf{A}_2=\sigma^2_rg_{ib}\textbf{H}^H_{ib}\boldsymbol{\Psi}\boldsymbol{\Psi}^H\textbf{H}_{ib}+\sigma_b^2\textbf{I}_N.
\end{align}
In accordance with the generalized Rayleigh-Rize theorem, the optimal $\textbf{u}_b$ is given by the eigenvector corresponding to the largest eigenvalue of $\textbf{A}_2^{-1}\textbf{A}_1$.


\subsection{Optimization of the IRS  phase shift matrix $\boldsymbol{\Psi}$}

In the previous sections, the PA factors $\beta$ and $l$, transmit beamforming $\textbf{v}$, and receive beamforming $\textbf{u}_b$ have been optimized. In this section, we turn our focus to the optimization of the IRS phase shift matrix $\boldsymbol{\Psi}$. In what follows, two strategies for optimizing $\boldsymbol{\Psi}$ by fixing the variables $\beta$, $l$, $\textbf{v}$, and $\textbf{u}_b$ will be proposed.

\subsubsection{Max-SR-SS algorithm}
First, we transform the power constraint (\ref{P_r}) into a constraint on $\boldsymbol{\Psi}$. Based on the fact that $\text{diag}\{\textbf{p}\}\textbf{q}=\text{diag}\{\textbf{q}\}\textbf{p}$ for $\forall\textbf{p}, \textbf{q} \in \mathbb{C}^{M\times 1}$, (\ref{P_r}) can be re-arranged as follows
\begin{align}\label{phi_P0}
P_r&=\text{Tr}\left(\boldsymbol{\Psi}(g_{ai}\beta l P\textbf{H}_{ai}\textbf{v}\textbf{v}^H\textbf{H}_{ai}^H+
\sigma^2_r\textbf{I}_M)\boldsymbol{\Psi}^H\right)\nonumber\\
&=\boldsymbol{\psi}^T(g_{ai}\beta l P\text{diag}\{\textbf{v}^H\textbf{H}_{ai}^H\}\text{diag}\{\textbf{H}_{ai}\textbf{v}\}
+\sigma^2_r\textbf{I}_M)\boldsymbol{\psi}^*\nonumber\\
&\leq (1-l)P.
\end{align}

Given that the inverse operation in (\ref{y_e3}), it is difficult to tackle the optimization problem (\ref{p0}) directly. Hence, to transform $\widetilde{R}_e$ in (\ref{y_e3}) into an tractable form, let us define
\begin{subequations}
\begin{align}
&\textbf{H}_1=\sigma_r^2g_{ie}\text{diag}\{\textbf{h}_{ie}\}\text{diag}\{\textbf{h}^H_{ie}\},\\
&\textbf{H}_2=(1-\beta)lPg_{ae}\textbf{H}^H_{ae}\textbf{T}_{AN}\textbf{T}^H_{AN}\textbf{H}_{ae}+\sigma_e^2\textbf{I}_{N_e},\\
&\textbf{H}_3=\sqrt{\beta l P g_{aie}}\textbf{H}^H_{ie}\text{diag}\{\textbf{H}_{ai}\textbf{v}\},\\
&\textbf{e}=\sqrt{\beta l Pg_{ae}}\textbf{H}^H_{ae}\textbf{v}.
\end{align}
\end{subequations}
Then, we introduce a slack variable $t$, which meets
\begin{align}
t\geq& (\textbf{H}_3\boldsymbol{\psi}+\textbf{e})^H(\boldsymbol{\psi}^H\textbf{H}_1\boldsymbol{\psi}\textbf{h}_{ei}\textbf{h}^H_{ei}+\textbf{H}_2)^{-1}
(\textbf{H}_3\boldsymbol{\psi}+\textbf{e}).
\end{align}
In accordance with the nature of Schur complement, we can obtain
\begin{align}\label{S}
\textbf{S}(\boldsymbol{\psi}, t)=    &\left[
        {\begin{array}{cc}
\boldsymbol{\psi}^H\textbf{H}_1\boldsymbol{\psi}\textbf{h}_{ei}\textbf{h}^H_{ei}+\textbf{H}_2 &
           \textbf{H}_3\boldsymbol{\psi}+\textbf{e}\\
        \boldsymbol{\psi}^H\textbf{H}^H_3+\textbf{e}^H &
            t
        \end{array}} \right]
\succeq \textbf{0}.
\end{align}
According to the first-order Taylor approximation of $\boldsymbol{\psi}^H\textbf{H}_1\boldsymbol{\psi}$ at feasible point $\bar{\boldsymbol{\psi}}$, we have
$\boldsymbol{\psi}^H\textbf{H}_1\boldsymbol{\psi}\geq 2\Re\{\boldsymbol{\psi}^H\textbf{H}_1\bar{\boldsymbol{\psi}}\}-\bar{\boldsymbol{\psi}}^H\textbf{H}_1\bar{\boldsymbol{\psi}}.$
Then, (\ref{S}) can be rewritten as
\begin{align}\label{S1}
&\textbf{S}(\boldsymbol{\psi}, t)\succeq\nonumber\\
    &\left[
        {\begin{array}{cc}
(2\Re\{\boldsymbol{\psi}^H\textbf{H}_1\bar{\boldsymbol{\psi}}\}-\bar{\boldsymbol{\psi}}^H\textbf{H}_1\bar{\boldsymbol{\psi}})\textbf{h}_{ei}\textbf{h}^H_{ei}+\textbf{H}_2 &
           \textbf{H}_3\boldsymbol{\psi}+\textbf{e}\\
        \boldsymbol{\psi}^H\textbf{H}^H_3+\textbf{e}^H &
            t
        \end{array}} \right]
\succeq \textbf{0}.
\end{align}
At this point, the optimization problem with respect to $\boldsymbol{\Psi}$ can be recast as
\begin{subequations}\label{Phi1}
\begin{align}
&\max \limits_{\boldsymbol{\psi}, t}
~~R_b-\text{log}_2(1+t),\\
&~~\text{s.t.} ~~~|\boldsymbol{\psi}(m)|\leq {\psi}^{\text{max}},~ (\ref{phi_P0}),~ (\ref{S1}).
\end{align}
\end{subequations}
The objective function of the problem (\ref{Phi1}) is the difference of two logarithmic functions and is non-convex. To address this problem, let us define
\begin{subequations}
\begin{align}
&\textbf{a}^H=\sqrt{\beta lPg_{aib}}\textbf{u}_b^H\textbf{H}^H_{ib}\text{diag}\{\textbf{H}_{ai}\textbf{v}\},\\
&b_1=\sqrt{\beta lPg_{ab}}\textbf{u}_b^H\textbf{H}^H_{ab}\textbf{v},\\
&\textbf{C}_1=\sigma_r\sqrt{g_{ib}}\text{diag}\{\textbf{u}_b^H\textbf{H}_{ib}^H\}.
\end{align}
\end{subequations}
Then, we have
\begin{align}
R_b&=\text{log}_2\left(1+\frac{|\textbf{a}^H\boldsymbol{\psi}+b_1|^2}{\|\textbf{C}_1\boldsymbol{\psi}\|^2+\sigma_b^2}\right)\nonumber\\
&=\text{log}_2(\boldsymbol{\psi}^H(\underbrace{\textbf{a}\textbf{a}^H+\textbf{C}_1^H\textbf{C}_1}_{\textbf{E}})\boldsymbol{\psi}+2\Re\{b_1^*\textbf{a}^H\boldsymbol{\psi}\}+|b_1|^2+\sigma_b^2)\nonumber\\
&~~~-\text{log}_2(1+\|\textbf{C}_1\boldsymbol{\psi}\|^2/\sigma_b^2)-\text{log}_2(\sigma_b^2).
\end{align}
Based on the first-order Taylor approximation of $\boldsymbol{\psi}^H\textbf{E}\boldsymbol{\psi}$, i.e., $\boldsymbol{\psi}^H\textbf{E}\boldsymbol{\psi}\geq 2\Re\{\boldsymbol{\psi}^H\textbf{E}\bar{\boldsymbol{\psi}}\}-\bar{\boldsymbol{\psi}}^H\textbf{E}\bar{\boldsymbol{\psi}}$ and the result in \cite{Nasir2017Secrecy},  for fixed points $\bar{e}_1$,
\begin{align}\label{e2}
&-\text{In}(1+e_1)\geq -\text{In}(1+\bar{e}_1)-\frac{1+e_1}{1+\bar{e}_1}+1,
\end{align}
after neglecting the constant entries, (\ref{Phi1}) can be recast as
\begin{subequations}\label{psi00}
\begin{align}
&\max \limits_{\boldsymbol{\psi}, t}
~~\text{In}(2\Re\{\boldsymbol{\psi}^H\textbf{E}\bar{\boldsymbol{\psi}}\}-\bar{\boldsymbol{\psi}}^H\textbf{E}\bar{\boldsymbol{\psi}}+2\Re\{b_1^*\textbf{a}^H\boldsymbol{\psi}\}+|b_1|^2+\sigma_b^2)\nonumber\\
&~~~~~~~~~-\frac{\|\textbf{C}_1\boldsymbol{\psi}\|^2}{\sigma_b^2}/\big(1+{\|\textbf{C}_1\bar{\boldsymbol{\psi}}\|^2}/{\sigma_b^2}\big)-\frac{t}{1+\bar{t}}\\
&~~\text{s.t.} ~~~|\boldsymbol{\psi}(m)|\leq {\psi}^{\text{max}},~ (\ref{phi_P0}),~ (\ref{S1}),
\end{align}
\end{subequations}
where $\bar{t}$ stands for the value obtained at the previous iteration of $t$. It is noted that the problem (\ref{psi00}) is convex, which can be derived directly with convex optimizing toolbox.

\subsubsection{Max-SR-EM algorithm}

In the previous subsection, a Max-SR-SS scheme has been proposed to optimize the IRS phase shift matrix $\boldsymbol{\Psi}$, which has a high computational complexity. To reduce the complexity, a Max-SR-EM scheme with lower complexity is proposed in this section. Given that $\boldsymbol{\Psi}$ consists of amplitude and phase, we will derive $\boldsymbol{\Psi}$ by solving for them separately in the following.

Firstly, the derivation of the magnitude is taken into account. For the sake of derivation, we assume that $|\boldsymbol{\Psi}(m,m)|\leq {\psi}^{\text{max}}$ in (\ref{p0}) always holds and the amplitude of each IRS phase shift elements is the same, noted as $|\boldsymbol{\Psi}(m,m)|=\alpha_m=\alpha$, and $\boldsymbol{\Theta}=\text{diag}\{e^{j{\phi}_1}, \cdots, e^{j\phi_m}, \cdots, e^{j\phi_M}\}\in \mathbb{C}^{M\times M}$. Then, we have $\boldsymbol{\Psi}=\alpha \boldsymbol{\Theta}$. Based on the IRS power constraint (\ref{P_r})
and the fact that it is optimal when taking the equivalent value, i.e.,
\begin{align}
\text{Tr}\left(\alpha \boldsymbol{\Theta}(g_{ai}\beta l P\textbf{H}_{ai}\textbf{v}\textbf{v}^H\textbf{H}_{ai}^H+
\sigma^2_r\textbf{I}_M)\alpha \boldsymbol{\Theta}^H\right)= (1-l)P,
\end{align}
which yields the amplitude
\begin{align}\label{alpha0}
\alpha&=\sqrt{\frac{(1-l)P}{\text{Tr}\left(\boldsymbol{\Theta}(g_{ai}\beta l P\textbf{H}_{ai}\textbf{v}\textbf{v}^H\textbf{H}_{ai}^H+\sigma^2_r\textbf{I}_M)\boldsymbol{\Theta}^H\right)}}\nonumber\\
&=\sqrt{\frac{(1-l)P}{\text{Tr}\left(g_{ai}\beta l P\textbf{H}_{ai}\textbf{v}\textbf{v}^H\textbf{H}_{ai}^H+\sigma^2_r\textbf{I}_M\right)}}.
\end{align}

In the following, we focus on finding the phase matrix $\boldsymbol{\Theta}$. Let us define
\begin{subequations}
\begin{align}
&\boldsymbol{\theta}=[e^{j{\phi}_1}, \cdots, e^{j\phi_m}, \cdots, e^{j\phi_M}]^T, ~\boldsymbol{\phi}=[\boldsymbol{\theta};1],\\
&\textbf{H}_{e1}=\text{blkdiag}\big\{\alpha^2\textbf{H}_1, 0\big\},~\textbf{H}_{e2}=[\alpha\textbf{H}_3~~\textbf{e}],\\
&\textbf{h}^H_{b}=[\alpha\sqrt{\beta l Pg_{aib}}\textbf{u}_b^H\textbf{H}_{ib}^H\text{diag}\{\textbf{H}_{ai}\textbf{v}\}~~\sqrt{\beta l Pg_{ab}}\textbf{u}_b^H\textbf{H}_{ab}\textbf{v}],\\
&\textbf{H}_{b}=\text{blkdiag}\big\{\alpha\sigma_r\text{diag}\{\sqrt{g_{ib}}\textbf{u}_b^H\textbf{H}_{ib}^H\}, 0\big\}.
\end{align}
\end{subequations}
Then, (\ref{y_b3}) and (\ref{y_e3}) can be rewritten as
\begin{align}\label{Rb5}
R_b&=\text{log}_2\left(1+\frac{|\textbf{h}^H_{b}\boldsymbol{\phi}|^2}{\|\textbf{H}_{b}\boldsymbol{\phi}\|^2+\sigma_b^2}\right),
\end{align}
and
\begin{align}
\widetilde{R}_e=\text{log}_2\Big(1+\text{Tr}\Big[\frac{\textbf{H}_{e2}\boldsymbol{\phi}\boldsymbol{\phi}^H\textbf{H}_{e2}^H}
{\boldsymbol{\phi}^H\textbf{H}_{e1}\boldsymbol{\phi}\textbf{h}_{ei}\textbf{h}^H_{ei}+\textbf{H}_2}\Big]\Big),
\end{align}
respectively.

Next, we perform a transformation of $R_b$. By (\ref{Rb5}) and the fact that for fixed points $\bar{e}_2$ and $\bar{e}_3$,
\begin{align}\label{log1}
&\text{In}\left(1+\frac{|e_2|^2}{e_3}\right)\geq \text{In}\left(1+\frac{|\bar{e}_2|^2}{\bar{e}_3}\right)-\frac{|\bar{e}_2|^2}{\bar{e}_3}+
\frac{2\Re\{\bar{e}_2e_2\}}{\bar{e}_3}-\nonumber\\
&~~~~~~~~~~~~~~~~~~~~~\frac{|\bar{e}_2|^2}{\bar{e}_3(\bar{e}_3+|\bar{e}_2|^2)}\left(e_3+|e_2|^2\right),
\end{align}
one obtains
\begin{align}\label{tau}
&R_b\cdot\text{In}2=\text{In}\left(1+\frac{|\textbf{h}^H_{b}\boldsymbol{\phi}|^2}{\|\textbf{H}_{b}\boldsymbol{\phi}\|^2+\sigma_b^2}\right)\nonumber\\
&=\text{In}\left(1+\frac{|\textbf{h}^H_{b}\bar{\boldsymbol{\phi}}|^2}{\tau}\right)-\frac{|\textbf{h}^H_{b}\bar{\boldsymbol{\phi}}|^2}{\tau}+
\frac{2\Re\{\boldsymbol{\phi}^H\textbf{h}_b\textbf{h}_b^H\bar{\boldsymbol{\phi}}\}}{\tau}+G,
\end{align}
where $\tau=\|\textbf{H}_b\bar{\boldsymbol{\phi}}\|^2+\sigma_b^2$ and
\begin{align}\label{G}
G=-\boldsymbol{\phi}^H\Big(\underbrace{\frac{|\textbf{h}_b^H\bar{\boldsymbol{\phi}}|^2}
{\tau(\tau+\|\textbf{H}_b\bar{\boldsymbol{\phi}}\|^2)}(\textbf{H}_b^H\textbf{H}_b+\textbf{h}_b\textbf{h}_b^H)}_{\widetilde{\textbf{M}}}\Big)\boldsymbol{\phi}^H.
\end{align}

With the majorization-minimization (MM) algorithm in \cite{Sun2017MM}, i.e.,
\begin{align}\label{lambda0}
-\textbf{x}^H\textbf{Y}\textbf{x}\geq -\textbf{x}^H\textbf{Z}\textbf{x}-2\Re\{\textbf{x}^H(\textbf{Y}-\textbf{Z})\overline{\textbf{x}}\}-\overline{\textbf{x}}^H(\textbf{Z}-\textbf{Y})\overline{\textbf{x}},
\end{align}
where $\textbf{Z}=\lambda_{\max}(\textbf{Y})\textbf{I}$, (\ref{G}) can be recast as
\begin{align}\label{phi3}
&-\boldsymbol{\phi}^H\widetilde{\textbf{M}}\boldsymbol{\phi}\geq -\boldsymbol{\phi}^H\lambda_{\max}(\widetilde{\textbf{M}})\textbf{I}_{M+1}\boldsymbol{\phi}-2\Re\{\boldsymbol{\phi}^H(\widetilde{\textbf{M}}-\nonumber\\
&\lambda_{\max}(\widetilde{\textbf{M}})\textbf{I}_{M+1})\bar{\boldsymbol{\phi}}\}-\bar{\boldsymbol{\phi}}^H(\lambda_{\max}(\widetilde{\textbf{M}})\textbf{I}_{M+1}-
\widetilde{\textbf{M}})\bar{\boldsymbol{\phi}}.
\end{align}

Next, we transform $\widetilde{R}_e$ in (\ref{y_e3}) into a form that is tractable to solving. Based on the fact that for $\forall~\textbf{X}\in \mathbb{C}^{X\times Y}$ and $\textbf{Y}\in \mathbb{C}^{Y\times X}$, one has
\begin{align}
|\textbf{I}_X+\textbf{X}\textbf{Y}|=|\textbf{I}_Y+\textbf{Y}\textbf{X}|.
\end{align}
Then, we have
\begin{align}\label{term0}
&\text{log}_2\Big(1+\text{Tr}\Big[\frac{\textbf{H}_{e2}\boldsymbol{\phi}\boldsymbol{\phi}^H\textbf{H}_{e2}^H}
{\boldsymbol{\phi}^H\textbf{H}_{e1}\boldsymbol{\phi}\textbf{h}_{ei}\textbf{h}^H_{ei}+\textbf{H}_2}\Big]\Big)\cdot\text{In}2\nonumber\\
&=\text{In}|1+\boldsymbol{\phi}^H\textbf{H}_{e2}^H(\boldsymbol{\phi}^H\textbf{H}_{e1}\boldsymbol{\phi}\textbf{h}_{ei}\textbf{h}^H_{ei}+\textbf{H}_2)^{-1}\textbf{H}_{e2}\boldsymbol{\phi}|\nonumber\\
&=\text{In}|\textbf{I}_{N_e}+(\boldsymbol{\phi}^H\textbf{H}_{e1}\boldsymbol{\phi}\textbf{h}_{ei}\textbf{h}^H_{ei}+\textbf{H}_2)^{-1}
\textbf{H}_{e2}\boldsymbol{\phi}\boldsymbol{\phi}^H\textbf{H}_{e2}^H|\nonumber\\
&=\text{In}|\boldsymbol{\phi}^H\textbf{H}_{e1}\boldsymbol{\phi}\textbf{h}_{ei}\textbf{h}^H_{ei}+\textbf{H}_2+\textbf{H}_{e2}\boldsymbol{\phi}\boldsymbol{\phi}^H\textbf{H}_{e2}^H|-\nonumber\\
&~~~~\text{In}|\boldsymbol{\phi}^H\textbf{H}_{e1}\boldsymbol{\phi}\textbf{h}_{ei}\textbf{h}^H_{ei}+\textbf{H}_2|\nonumber\\
&=\text{In}|\boldsymbol{\phi}^H\textbf{H}_{e1}\boldsymbol{\phi}\textbf{h}_{ei}\textbf{h}^H_{ei}+\textbf{H}_2+\textbf{H}_{e2}\boldsymbol{\phi}\boldsymbol{\phi}^H\textbf{H}_{e2}^H|-\nonumber\\
&~~~~\text{In}(|\boldsymbol{\phi}^H\textbf{H}_{e1}\boldsymbol{\phi}\textbf{h}_{ei}\textbf{h}^H_{ei}\textbf{H}_2^{-1}+\textbf{I}_{N_e}|\textbf{H}_2|)\nonumber\\
&=\text{In}|\underbrace{\boldsymbol{\phi}^H\textbf{H}_{e1}\boldsymbol{\phi}\textbf{h}_{ei}\textbf{h}^H_{ei}+\textbf{H}_2+\textbf{H}_{e2}\boldsymbol{\phi}\boldsymbol{\phi}^H\textbf{H}_{e2}^H}
_{\textbf{J}}|-\nonumber\\
&~~~~\text{In}(1+\underbrace{\boldsymbol{\phi}^H\textbf{H}_{e1}\boldsymbol{\phi}\textbf{h}^H_{ei}\textbf{H}_2^{-1}\textbf{h}_{ei}}_{\eta})-\text{In}|\textbf{H}_2|.
\end{align}

To simplify the first term of (\ref{term0}), based on
\begin{align}\label{abs1}
\text{In}|\textbf{X}|\leq\text{In}|\bar{\textbf{X}}|+\text{Tr}[\bar{\textbf{X}}^{-1}(\textbf{X}-\bar{\textbf{X}})],
\end{align}
we have
\begin{align}
-\text{In}|\textbf{J}|&\geq -\text{In}|\bar{\textbf{J}}|-\text{Tr}[\bar{\textbf{J}}^{-1}(\textbf{J}-\bar{\textbf{J}})]\nonumber\\
&=-\text{In}|\bar{\textbf{J}}|+\text{Tr}[\bar{\textbf{J}}^{-1}\bar{\textbf{J}}]-\text{Tr}[\bar{\textbf{J}}^{-1}\textbf{J}]\nonumber\\
&=-\text{In}|\bar{\textbf{J}}|+\text{Tr}[\bar{\textbf{J}}^{-1}\bar{\textbf{J}}]-\boldsymbol{\phi}^H\textbf{H}_{e1}\boldsymbol{\phi}\text{Tr}
[\bar{\textbf{J}}^{-1}\textbf{h}_{ei}\textbf{h}^H_{ei}]-\nonumber\\
&~~~~\text{Tr}[\bar{\textbf{J}}^{-1}\textbf{H}_2]-\boldsymbol{\phi}^H\textbf{H}_{e2}^H\bar{\textbf{J}}^{-1}\textbf{H}_{e2}\boldsymbol{\phi},\nonumber\\
&=-\text{In}|\bar{\textbf{J}}|+\text{Tr}[\bar{\textbf{J}}^{-1}\bar{\textbf{J}}]-\text{Tr}[\bar{\textbf{J}}^{-1}\textbf{H}_2]-\nonumber\\
&~~~~\boldsymbol{\phi}^H(\underbrace{\textbf{H}_{e1}\text{Tr}[\bar{\textbf{J}}^{-1}\textbf{h}_{ei}\textbf{h}^H_{ei}]+\textbf{H}_{e2}^H\bar{\textbf{J}}^{-1}\textbf{H}_{e2}}_{\textbf{K}})
\boldsymbol{\phi},
\end{align}
where $\bar{\textbf{J}}$ means the solution obtained at the previous iteration of $\textbf{J}$. By utilizing (\ref{lambda0}), one has
\begin{align}\label{term1}
-\boldsymbol{\phi}^H\textbf{K}\boldsymbol{\phi}&\geq -\boldsymbol{\phi}^H\lambda_{\text{max}}(\textbf{K})\textbf{I}_{M+1}\boldsymbol{\phi}-2\Re\{\boldsymbol{\phi}^H(\textbf{K}-\nonumber\\
&\lambda_{\text{max}}(\textbf{K})\textbf{I}_{M+1})\bar{\boldsymbol{\phi}}\}-\bar{\boldsymbol{\phi}}^H(\lambda_{\text{max}}(\textbf{K})\textbf{I}_{M+1}-\textbf{K})\bar{\boldsymbol{\phi}}.
\end{align}
To make the second term of (\ref{term0}) tractable, according to (\ref{e2}), we can obtain
\begin{align}
-\text{In}(1+\eta)\geq &-\text{In}(1+\bar{\eta})-\frac{1+\boldsymbol{\phi}^H\textbf{H}_{e1}\boldsymbol{\phi}\textbf{h}^H_{ei}\textbf{H}_2^{-1}\textbf{h}_{ei}}{1+\bar{\eta}}+1,
\end{align}
where $\bar{\eta}$ is the solution obtained at the previous iteration. Based on the first-order Taylor series expansion, we have
\begin{align}\label{term2}
\frac{\boldsymbol{\phi}^H\textbf{H}_{e1}\boldsymbol{\phi}\textbf{h}^H_{ei}\textbf{H}_2^{-1}\textbf{h}_{ei}}{1+\bar{\eta}}\geq & 2\Re\{\boldsymbol{\phi}^H\frac{\textbf{H}_{e1}(\textbf{h}^H_{ei}\textbf{H}_2^{-1}\textbf{h}_{ei})}{1+\bar{\eta}}\bar{\boldsymbol{\phi}}\}-\nonumber\\
& \bar{\boldsymbol{\phi}}^H\frac{\textbf{H}_{e1}(\textbf{h}^H_{ei}\textbf{H}_2^{-1}\textbf{h}_{ei})}{1+\bar{\eta}}\bar{\boldsymbol{\phi}}.
\end{align}

At this point, combined with (\ref{tau}), (\ref{phi3}), (\ref{term1}), and (\ref{term2}), after neglecting the constant term, the optimization problem with respect to $\boldsymbol{\phi}$ can be recast as
\begin{subequations}\label{}
\begin{align}
&\max \limits_{\boldsymbol{\phi}}
~~2\Re\{\boldsymbol{\phi}^H\textbf{g}\}\\
&~~\text{s.t.} ~~~|\boldsymbol{\phi}(m)|=1, m=1,\cdots M, \\
&~~~~~~~~~\boldsymbol{\phi}(M+1)=1,
\end{align}
\end{subequations}
where
\begin{align}{}
\textbf{g}&=\Big[\frac{\textbf{h}_b\textbf{h}_b^H}{\tau}-(\widetilde{\textbf{M}}-\lambda_{\max}(\widetilde{\textbf{M}})\textbf{I}_{M+1})-
(\textbf{K}-\lambda_{\text{max}}(\textbf{K})\textbf{I}_{M+1})\nonumber\\
&~~~+\frac{\textbf{H}_{e1}(\textbf{h}^H_{ei}\textbf{H}_2^{-1}\textbf{h}_{ei})}{1+\bar{\eta}}\Big]\bar{\boldsymbol{\phi}}.
\end{align}
Then, the optimal solution of $\boldsymbol{\theta}$ can be obtain directly by
\begin{align}\label{theta0}
\boldsymbol{\theta}^{\text{opt}}=\boldsymbol{\phi}^{\text{opt}}(1:M)=e^{j\text{arg}(\textbf{g}(1:M))}.
\end{align}

\subsection{Overall scheme and complexity analysis}

Up to now, we have completed the derivation of the PA factors $\beta$ and $l$, transmit beamforming $\textbf{v}$, receive beamforming $\textbf{u}_b$, and IRS phase shift matrix $\boldsymbol{\Psi}$. To make the process of this scheme clearer, we summarize the entire proposed schemes below.

The iterative idea of the proposed Max-SR-SS scheme is as follows: (1) the PA factors $\beta$ and $l$, transmit beamforming $\textbf{v}$, receive beamforming $\textbf{u}_b$, and IRS phase shift matrix $\boldsymbol{\Psi}$ are initialized to feasible solutions; (2) given $l$, $\textbf{v}$, $\textbf{u}_b$, and $\boldsymbol{\Psi}$, based on Algorithm 1 to update $\beta$; (3) fixed $\beta$, $\textbf{v}$, $\textbf{u}_b$, and $\boldsymbol{\Psi}$, solve (\ref{L00}) to update $l$; (4) given $\beta$, $l$, $\textbf{u}_b$, and $\boldsymbol{\Psi}$, solve (\ref{v33}) to obtain $\textbf{v}$; (5) fixed $\beta$, $l$, $\textbf{v}$, and $\boldsymbol{\Psi}$, solve (\ref{u_b}) to yield $\textbf{u}_b$; (6) given $\beta$, $l$, $\textbf{v}$, and $\textbf{u}_b$, solve (\ref{psi00}) to yield $\boldsymbol{\psi}$, and $\boldsymbol{\Psi}=\text{diag}\{\boldsymbol{\psi}\}$. The five variables are updated alternately until the termination condition is realized, i.e., $|R_s^{(k)}-R_s^{(k-1)}|\leq\epsilon$, where $k$ and $\epsilon$ refer to the iteration number and convergence accuracy, respectively.

The overall procedure of the proposed Max-SR-EM scheme is listed below: (1) the PA factors $\beta$ and $l$, transmit beamforming $\textbf{v}$, receive beamforming $\textbf{u}_b$, and IRS phase shift matrix $\boldsymbol{\Psi}$ are initialized to feasible solutions; (2) given $l$, $\textbf{v}$, $\textbf{u}_b$, and $\boldsymbol{\Psi}$, $\beta$ is computed by the Algorithm 1; (3) fixed $\beta$, $\textbf{v}$, $\textbf{u}_b$, and $\boldsymbol{\Psi}$, $l$ is updated by (\ref{L00}); (4) given $\beta$, $l$, $\textbf{u}_b$, and $\boldsymbol{\Psi}$, $\textbf{v}$ is updated by (\ref{v33}); (5) fixing $\beta$, $l$, $\textbf{v}$, and $\boldsymbol{\Psi}$, $\textbf{u}_b$ is derived via the generalized Rayleigh-Ritz theorem; (6) given $\beta$, $l$, $\textbf{v}$, and $\textbf{u}_b$, solve (\ref{alpha0}) to obtain $\alpha$, solve (\ref{theta0}) to find $\boldsymbol{\theta}$, and $\boldsymbol{\Psi}=\alpha\text{diag}\{\boldsymbol{\theta}\}$. The alternating iteration is repeated until the termination condition is met.

Due to the fact that the obtained solutions in Max-SR-SS and Max-SR-EM schemes are sub-optimal, and the objective value sequence $\{R_s(\beta^{(k)}, l^{(k)}, \textbf{v}^{(k)}, \textbf{u}_b^{(k)}, \boldsymbol{\Psi}^{(k)})\}$ obtained in each iteration of the alternate optimization method is non-decreasing. Specifically, it follows
\begin{align}
&R_s\left(\beta^{(k)}, l^{(k)}, \textbf{v}^{(k)}, \textbf{u}_b^{(k)}, \boldsymbol{\Psi}^{(k)}\right)\nonumber\\
&\overset{(a)}{\leq}R_s\left(\beta^{(k+1)}, l^{(k)}, \textbf{v}^{(k)}, \textbf{u}_b^{(k)}, \boldsymbol{\Psi}^{(k)}\right)\nonumber\\
&\overset{(b)}{\leq}R_s\left(\beta^{(k+1)}, l^{(k+1)}, \textbf{v}^{(k)}, \textbf{u}_b^{(k)}, \boldsymbol{\Psi}^{(k)}\right)\nonumber\\
&\overset{(c)}{\leq}R_s\left(\beta^{(k+1)}, l^{(k+1)}, \textbf{v}^{(k+1)}, \textbf{u}_b^{(k)}, \boldsymbol{\Psi}^{(k)}\right)\nonumber\\
&\overset{(d)}{\leq}R_s\left(\beta^{(k+1)}, l^{(k+1)}, \textbf{v}^{(k+1)}, \textbf{u}_b^{(k+1)}, \boldsymbol{\Psi}^{(k)}\right)\nonumber\\
&\overset{(e)}{\leq}R_s\left(\beta^{(k+1)}, l^{(k+1)}, \textbf{v}^{(k+1)}, \textbf{u}_b^{(k+1)}, \boldsymbol{\Psi}^{(k+1)}\right),
\end{align}
where $(a)$, $(b)$, $(c)$, $(d)$ and $(e)$ are due to the update of $\beta$, $l$, $\textbf{v}$, $\textbf{u}_b$, and $\boldsymbol{\Psi}$, respectively. Moreover, $R_s(\beta^{(k)}, l^{(k)}, \textbf{v}^{(k)}, \textbf{u}_b^{(k)}, \boldsymbol{\Psi}^{(k)})$ has a finite upper bound since the limited power constraint. Therefore, the convergence of the proposed three schemes can be guaranteed.

Next, we calculate the computational complexity of the two proposed schemes.

1) For the Max-SR-SS scheme, the overall computational complexity is $C_{SS}=\mathcal {O}\{L_{SS}[(\sqrt{5}(N_eM^3+N_e M^2)+N^3M^2+NM)1/{\xi}+(N_b+N_e)M^2+N_b^3]\}$
float-point operations (FLOPs), where $L_{SS}$ refers to the maximum number of alternating iterations, $\xi$ stands for the given accuracy tolerance of CVX.

2) For the Max-SR-EM scheme, the whole computational complexity is $C_{EM}=\mathcal {O}\{L_{EM}[(N^3M^2+NM)1/{\xi}+(N_b+2N_e)M^2+N_eM+N_b^3]\}$
FLOPs, where $L_{FS}$ represents the maximum number of alternating iterations.

It is not difficult to find that the computational complexity of the two proposed schemes can be listed in decreasing order as $C_{SS}>C_{EM}$.

\section{Proposed Max-SR-AO scheme}\label{s3}
In this section, we consider a special situation of problem (\ref{p0}), i.e., both of Bob and Eve are equipped with single antenna. At this point, the channels $\textbf{H}_{ab}$, $\textbf{H}_{ae}$, $\textbf{H}_{ib}$, $\textbf{H}_{ie}$ are degenerated to $\textbf{h}_{ab}\in \mathbb{C}^{N\times 1}$, $\textbf{h}_{ae}\in \mathbb{C}^{N\times 1}$, $\textbf{h}_{ib}\in \mathbb{C}^{M\times 1}$, $\textbf{h}_{ie}\in \mathbb{C}^{M\times 1}$, respectively, and the receive beamforming is not done. Then, the receive signal (\ref{y_b1}) and (\ref{y_e1}) can be degenerated to
\begin{align}\label{yy_b2}
y_b&=\sqrt{\beta l P}
\left(\sqrt{g_{ab}}\textbf{h}^H_{ab}+\sqrt{g_{aib}}\textbf{h}^H_{ib}\boldsymbol{\Psi}\textbf{H}_{ai}\right)\textbf{v}x+
\sqrt{g_{ib}}\textbf{h}^H_{ib}\boldsymbol{\Psi}\textbf{n}_r\nonumber\\
&~~~+n_b,
\end{align}
and
\begin{align}\label{yy_e2}
y_e&=\sqrt{\beta l P}\left(\sqrt{g_{ae}}\textbf{h}^H_{ae}+
\sqrt{g_{aie}}\textbf{h}^H_{ie}\boldsymbol{\Psi}\textbf{H}_{ai}\right)\textbf{v}x+\nonumber\\
&~~~\sqrt{(1-\beta) l P}\sqrt{g_{ae}}\textbf{h}^H_{ae}\textbf{T}_{AN}\textbf{z}+
\sqrt{g_{ie}}\textbf{h}^H_{ie}\boldsymbol{\Psi}\textbf{n}_r+n_e,
\end{align}
respectively. Correspondingly, the achievable rates at Bob and Eve are respectively given by
\begin{align}\label{yy_b3}
R_b=\text{log}_2\Big(1+\frac{\beta l P|\left(\sqrt{g_{ab}}\textbf{h}^H_{ab}+
\sqrt{g_{aib}}\textbf{h}^H_{ib}\boldsymbol{\Psi}\textbf{H}_{ai}\right)\textbf{v}|^2}
{\sigma^2_r\|\sqrt{g_{ib}}\textbf{h}^H_{ib}\boldsymbol{\Psi}\|^2+\sigma^2_b}\Big),
\end{align}
and
\begin{align}\label{yy_e3}
R_e&=\text{log}_2\Big(1+\nonumber\\
&\frac{\beta l P|\left(\sqrt{g_{ae}}\textbf{h}^H_{ae}+
\sqrt{g_{aie}}\textbf{h}^H_{ie}\boldsymbol{\Psi}\textbf{H}_{ai}\right)\textbf{v}|^2}
{(1-\beta) l P \|\sqrt{g_{ae}}\textbf{h}^H_{ae}\textbf{T}_{AN}\|^2+
\sigma^2_r\|\sqrt{g_{ie}}\textbf{h}^H_{ie}\boldsymbol{\Psi}\|^2+\sigma^2_e}\Big).
\end{align}
In the absence of receive beamforming, the optimization problem (\ref{p0}) can be recast as
\begin{subequations}\label{pp0}
\begin{align}
&\max \limits_{\beta, l, \textbf{v}, \boldsymbol{\Psi}}
~~R_s=R_b-R_e\\
&~~~\text{s.t.} ~~~~\textbf{v}^H\textbf{v}=1,~P_r\leq (1-l)P,\\
& ~~~~~~~~~~0<\beta\leq 1,~0<l< 1.\\
& ~~~~~~~~~~|\boldsymbol{\Psi}(m,m)|\leq {\psi}^{\text{max}}, m=1,\dots, M.
\end{align}
\end{subequations}
In what follows, the alternating iteration strategy is taken into account for solving the variables $\beta$, $l$, $\textbf{v}$, and $\boldsymbol{\Psi}$.

\subsection{Optimization of the PA factor $\beta$}
In this subsection, the beamforming vector $\textbf{v}$ and IRS phase shift matrix $\boldsymbol{\Psi}$ are given for the sake of simplicity.
Let us define
$D_b=P|(\sqrt{g_{ab}}\textbf{h}^H_{ab}+
\sqrt{g_{aib}}\textbf{h}^H_{ib}\boldsymbol{\Psi}\textbf{H}_{ai})\textbf{v}|^2,$
$D_e=P|(\sqrt{g_{ae}}\textbf{h}^H_{ae}+
\sqrt{g_{aie}}\textbf{h}^H_{ie}\boldsymbol{\Psi}\textbf{H}_{ai})\textbf{v}|^2,$
$E_b=\sigma^2_r\|\sqrt{g_{ib}}\textbf{h}^H_{ib}\boldsymbol{\Psi}\|^2+\sigma^2_b,$
$E_e=\sigma^2_r\|\sqrt{g_{ie}}\textbf{h}^H_{ie}\boldsymbol{\Psi}\|^2+\sigma^2_e,$
$F_e=P \|\sqrt{g_{ae}}\textbf{h}^H_{ae}\textbf{T}_{AN}\|^2.$
Then, (\ref{yy_b3}) and (\ref{yy_e3}) can be transformed into
\begin{align}
&R_b=\text{log}_2\left(\frac{\beta l D_b+E_b}{E_b}\right),
\end{align}
and
\begin{align}
&R_e=\text{log}_2\left(\frac{\beta l D_e+(1-\beta) l F_e+E_e}{(1-\beta) l F_e+E_e}\right),
\end{align}
respectively. The objective function of the optimization problem (\ref{pp0}) can be degenerated as
\begin{align}\label{R_s3}
&R_s=R_b-R_e\nonumber\\
&=\text{log}_2\left(\frac{(\beta l D_b+E_b)[(1-\beta) l F_e+E_e]}{\beta l D_e+(1-\beta) l F_e+E_e}\right)-\text{log}_2E_b\nonumber\\
&=\text{log}_2\frac{\beta(1-\beta)l^2D_bF_e+\beta l D_bE_e+(1-\beta)l E_bF_e+E_bE_e}{\beta l D_e+(1-\beta) l F_e+E_e}\nonumber\\
&~~-\text{log}_2E_b.
\end{align}
In what follows, we handle the optimization of the PA parameters $\beta$ and $l$ successively.

Given $l$, in accordance with (\ref{pp0}) and (\ref{R_s3}), the optimization problem with respect to $\beta$ can be simplified as follows
\begin{subequations}\label{}
\begin{align}
&\max \limits_{\beta}
~~\frac{1}{\beta (l D_e- l F_e)+l F_e+E_e}\Big(-\beta^2l^2D_bF_e+\nonumber\\
&~~~~\beta \left(l^2D_bF_e+l D_bE_e-l E_bF_e\right)+l E_bF_e+E_bE_e\Big)\\
&~~\text{s.t.} ~~~(\ref{beta1}), 0<\beta\leq 1,
\end{align}
\end{subequations}
which can be re-arrange as
\begin{subequations}\label{beta2}
\begin{align}
&\max \limits_{\beta}
~~\frac{-\beta^2A_3+\beta B_3+C_3}{\beta D_3+K_3}\label{object}\\
&~~\text{s.t.} ~~~\beta F_3\leq G_3, 0<\beta\leq 1,
\end{align}
\end{subequations}
where
$A_3=l^2D_bF_e, $
$B_3=l^2D_bF_e+l D_bE_e-l E_bF_e,$
$C_3=l E_bF_e+E_bE_e, $
$D_3=l D_e- l F_e,$
$K_3=l F_e+E_e, $
$F_3=l \text{Tr}\left(\boldsymbol{\Psi}(g_{ai} P\textbf{H}_{ai}\textbf{v}\textbf{v}^H\textbf{H}_{ai}^H)\boldsymbol{\Psi}^H\right),$
$G_3= (1-l)P-\text{Tr}(\sigma^2_r\boldsymbol{\Psi}\boldsymbol{\Psi}^H).$
It can be found that this problem is non-convex. Notice that this is a FP problem, and the denominator of (\ref{object}) is $\beta D_3+K_3=\beta l D_e+(1-\beta)lF_e+E_e>0$.
To transform (\ref{beta2}) into a convex optimization problem, based on the Dinkelbach's transform in \cite{Dinkelbach1967On}, we introduce a auxiliary parameter $\tau_1$ and recast the problem (\ref{beta2}) as follows
\begin{subequations}\label{A1}
\begin{align}
&\max \limits_{\beta, \tau_1}
~~{-\beta^2A_3+\beta B_3+C_3}-\tau_1(\beta D_3+K_3)\\
&~~\text{s.t.} ~~~\beta F_3\leq G_3, 0<\beta\leq 1.
\end{align}
\end{subequations}
The optimal solution can be obtained by taking the root of ${-\beta^2A_3+\beta B_3+C_3}-\tau_1(\beta D_3+K_3)=0$. At this point, the optimization problem (\ref{A1}) is convex, and we can address it by CVX directly.

\subsection{Optimization of the PA factor $l$}
Fixed$\beta$, $\textbf{v}$ and $\boldsymbol{\Psi}$, we transfer the focus to solving for $l$.
In accordance with (\ref{pp0}) and (\ref{R_s3}), by neglecting the constant terms, the optimization problem with respect to $l$ can be simplified as follows
\begin{subequations}\label{LL}
\begin{align}
&\max \limits_{l}
~~\frac{l^2\beta(1-\beta)D_bF_e+l(\beta D_bE_e+(1-\beta)E_bF_e)+E_bE_e}
{l(\beta D_e+(1-\beta)F_e)+E_e}\\
&~~\text{s.t.} ~~~(\ref{beta1}), 0<l< 1,
\end{align}
\end{subequations}
which yields
\begin{subequations}\label{L2}
\begin{align}
&\max \limits_{l}
~~\frac{l^2A_4+lB_4+C_4}{lD_4+K_4}\\
&~~\text{s.t.} ~~~l F_4\leq G_4, 0<l< 1,
\end{align}
\end{subequations}
where
$A_4=\beta(1-\beta)D_bF_e,$
$B_4=\beta D_bE_e+(1-\beta)E_bF_e,$
$C_4=E_bE_e,$
$D_4=\beta D_e+(1-\beta)F_e,$
$K_4=E_e,$
$F_4=\beta \text{Tr}\left(\boldsymbol{\Psi}(g_{ai} P\textbf{H}_{ai}\textbf{v}\textbf{v}^H\textbf{H}_{ai}^H)\boldsymbol{\Psi}^H\right)+P,$
$G_4=P-\text{Tr}(\sigma^2_r\boldsymbol{\Psi}\boldsymbol{\Psi}^H).$
It is noticed that $lD_4+K_4>0$, and this is a non-convex fractional optimization problem,  in accordance with the FP method, we introduce a auxiliary parameter $\tau_2$ and recast the problem (\ref{L2}) as
\begin{subequations}\label{SCA}
\begin{align}
&\max \limits_{l, \tau_2}
~~{l^2A_4+lB_4+C_4}-\tau_2(lD_4+K_4)\\
&~~\text{s.t.} ~~~l F_4\leq G_4, 0<l< 1,
\end{align}
\end{subequations}
The optimal solution to this problem is the root of ${l^2A_4+lB_4+C_4}-\tau_2(lD_4+K_4)=0$. However, the problem (\ref{SCA}) is still non-convex and requires further transformation.
With the first-order Taylor approximation of $l^2A_4$ at feasible point $\bar{l}$, i.e.,
$l^2A_4\geq 2\bar{l}A_4 l-\bar{l}^2A_4$, (\ref{SCA}) can be converted to
\begin{subequations}\label{L1}
\begin{align}
&\max \limits_{l, \tau_2}
~~{2\bar{l}A_4 l-\bar{l}^2A_4+lB_4+C_4}-\tau_2(lD_4+K_4)\\
&~~\text{s.t.} ~~~l F_4\leq G_4, 0<l< 1,
\end{align}
\end{subequations}
which is a convex optimization problem and can be addressed directly by the convex optimizing toolbox.

\subsection{Optimization of the beamforming vector $\textbf{v}$}
Given $\beta$, $l$, and $\boldsymbol{\Psi}$ with ignoring the constant term, (\ref{pp0}) can be reformulated as the optimization problem with respect to $\textbf{v}$ as follows
\begin{subequations}\label{vv1}
\begin{align}
&\max \limits_{\textbf{v}}
~~\frac{\textbf{v}^H\textbf{F}_1\textbf{v}}{\textbf{v}^H\textbf{F}_2\textbf{v}}\\
&~~\text{s.t.} ~~~\textbf{v}^H\textbf{v}=1, (\ref{P_r1}),
\end{align}
\end{subequations}
where
\begin{align}
\textbf{F}_1&=\beta l P\left(\sqrt{g_{ab}}\textbf{h}^H_{ab}+
\sqrt{g_{aib}}\textbf{h}^H_{ib}\boldsymbol{\Psi}\textbf{H}_{ai}\right)^H\big(\sqrt{g_{ab}}\textbf{h}^H_{ab}+\nonumber\\
&~~~\sqrt{g_{aib}}\textbf{h}^H_{ib}\boldsymbol{\Psi}\textbf{H}_{ai}\big)
+\left(\sigma^2_r\|\sqrt{g_{ib}}\textbf{h}^H_{ib}\boldsymbol{\Psi}\|^2+\sigma^2_b\right)\textbf{I}_N,\\
\textbf{F}_2&=\beta l P\left(\sqrt{g_{ae}}\textbf{h}^H_{ae}+
\sqrt{g_{aie}}\textbf{h}^H_{ie}\boldsymbol{\Psi}\textbf{H}_{ai}\right)^H\big(\sqrt{g_{ae}}\textbf{h}^H_{ae}+\nonumber\\
&~~~\sqrt{g_{aie}}\textbf{h}^H_{ie}\boldsymbol{\Psi}\textbf{H}_{ai}\big)
+\big((1-\beta) l P \|\sqrt{g_{ae}}\textbf{h}^H_{ae}\textbf{T}_{AN}\|^2+\nonumber\\
&~~~\sigma^2_r\|\sqrt{g_{ie}}\textbf{h}^H_{ie}\boldsymbol{\Psi}\|^2+\sigma^2_e\big)\textbf{I}_N.
\end{align}

Based on (\ref{T}) and relaxed the constraint $\textbf{v}^H\textbf{v}=1$ to $\textbf{v}^H\textbf{v}\leq1$, the problem (\ref{vv1}) can be recast as
\begin{subequations}\label{v3}
\begin{align}
&\max \limits_{\textbf{v}}
~~-\frac{\bar{\textbf{v}}^H\textbf{F}_1\bar{\textbf{v}}}{(\bar{\textbf{v}}^H\textbf{F}_2\bar{\textbf{v}})^2}
\textbf{v}^H\textbf{F}_2\textbf{v}+\frac{2\Re\{\bar{\textbf{v}}^H\textbf{F}_1\textbf{v}\}}
{\bar{\textbf{v}}^H\textbf{F}_2\bar{\textbf{v}}}\\
&~~\text{s.t.} ~~~\textbf{v}^H\textbf{v}\leq1, (\ref{P_r1}),
\end{align}
\end{subequations}
It can be found that this is a convex optimization problem that can be tackled directly with convex optimizing toolbox.


\subsection{Optimization of the IRS  phase shift matrix $\boldsymbol{\Psi}$}
In this subsection, we turn our target to optimize $\boldsymbol{\Psi}$ with given $\beta$, $l$, and $\textbf{v}$. For the sake of derivation, let us define
\begin{align}
&\widetilde{\boldsymbol{\psi}}=\left[ \begin{array}{*{20}{c}}
\boldsymbol{\psi}\\
1
\end{array}\right]^*_{(M+1)\times 1},\\
&\textbf{h}_{jj}=\left[ \begin{array}{*{20}{c}}
\sqrt{g_{aij}}\text{diag}\{\textbf{h}^H_{ij}\}\textbf{H}_{ai}\textbf{v}\\
\sqrt{g_{aj}}\textbf{h}^H_{aj}\textbf{v}
\end{array}\right]_{(M+1)\times 1}, j=b, e,\\
&\textbf{H}_{jj}=\left[ \begin{array}{*{20}{c}}
\sqrt{g_{ij}}\text{diag}\{\textbf{h}^H_{ij}\}\\
\textbf{0}^H
\end{array}\right]_{(M+1)\times M}, j=b, e.
\end{align}
Then, the achievable rates (\ref{yy_b3}) and (\ref{yy_e3}) can be rewritten as
\begin{align}\label{Rb1}
R_b=\text{log}_2\left(1+\frac{\beta l P|\widetilde{\boldsymbol{\psi}}^H\textbf{h}_{bb}|^2}
{\sigma^2_r\|\widetilde{\boldsymbol{\psi}}^H\textbf{H}_{bb}\|^2+\sigma^2_b}\right),
\end{align}
and
\begin{align}\label{Re1}
&R_e =\nonumber\\
&\text{log}_2\Big(1+
\frac{\beta l P|\widetilde{\boldsymbol{\psi}}^H\textbf{h}_{ee}|^2}
{\sigma^2_r\|\widetilde{\boldsymbol{\psi}}^H\textbf{H}_{ee}\|^2+(1-\beta) l P \|\sqrt{g_{ae}}\textbf{h}^H_{ae}\textbf{T}_{AN}\|^2+\sigma^2_e}\Big)\nonumber\\
&=\text{log}_2\Big(1+\frac{\beta l P|\widetilde{\boldsymbol{\psi}}^H\textbf{h}_{ee}|^2+
\sigma^2_r\|\widetilde{\boldsymbol{\psi}}^H\textbf{H}_{ee}\|^2}{(1-\beta) l P \|\sqrt{g_{ae}}\textbf{h}^H_{ae}\textbf{T}_{AN}\|^2+\sigma^2_e}\Big)-\nonumber\\
&~~~\text{log}_2\Big(1+\frac{\sigma^2_r\|\widetilde{\boldsymbol{\psi}}^H\textbf{H}_{ee}\|^2}{(1-\beta) l P \|\sqrt{g_{ae}}\textbf{h}^H_{ae}\textbf{T}_{AN}\|^2+\sigma^2_e}\Big),
\end{align}
respectively.

In addition, the power constraint (\ref{P_r}) can be re-arranged as follows
\begin{align}\label{phi_P}
P_r&=\text{Tr}\left(\boldsymbol{\Psi}(g_{ai}\beta l P\textbf{H}_{ai}\textbf{v}\textbf{v}^H\textbf{H}_{ai}^H+
\sigma^2_r\textbf{I}_M)\boldsymbol{\Psi}^H\right)\nonumber\\
&=\widetilde{\boldsymbol{\psi}}^H\text{blkdiag}\big\{g_{ai}\beta l P\text{diag}\{\textbf{v}^H\textbf{H}_{ai}^H\}\text{diag}\{\textbf{H}_{ai}\textbf{v}\}+\nonumber\\
&~~~~\sigma^2_r\textbf{I}_M, 0\big\}\widetilde{\boldsymbol{\psi}}\nonumber\\
&\leq (1-l)P.
\end{align}

At this point, the optimization problem with respect to $\boldsymbol{\Psi}$ is given by
\begin{subequations}\label{phi2}
\begin{align}
&\max \limits_{\widetilde{\boldsymbol{\psi}}}
~~\text{log}_2\Big(1+\frac{\beta l P|\widetilde{\boldsymbol{\psi}}^H\textbf{h}_{bb}|^2}
{\sigma^2_r\|\widetilde{\boldsymbol{\psi}}^H\textbf{H}_{bb}\|^2+\sigma^2_b}\Big)+\nonumber\\
&~~~~~~~~\text{log}_2\Big(1+\frac{\sigma^2_r\|\widetilde{\boldsymbol{\psi}}^H\textbf{H}_{ee}\|^2}{(1-\beta) l P \|\sqrt{g_{ae}}\textbf{h}^H_{ae}\textbf{T}_{AN}\|^2+\sigma^2_e}\Big)-\nonumber\\
&~~~~~~~~\text{log}_2\Big(1+\frac{\beta l P|\widetilde{\boldsymbol{\psi}}^H\textbf{h}_{ee}|^2+
\sigma^2_r\|\widetilde{\boldsymbol{\psi}}^H\textbf{H}_{ee}\|^2}{(1-\beta) l P \|\sqrt{g_{ae}}\textbf{h}^H_{ae}\textbf{T}_{AN}\|^2+\sigma^2_e}\Big)\\
&~~\text{s.t.} ~~~|\widetilde{\boldsymbol{\psi}}(m)|\leq {\psi}^{\text{max}}, ~\widetilde{\boldsymbol{\psi}}(m+1)=1,~ (\ref{phi_P}).
\end{align}
\end{subequations}
This problem is non-convex and further transformation is required. According to (\ref{log1}) and (\ref{e2}), by omitting the constant term, the optimization problem (\ref{phi2}) can be degenerated to
\begin{subequations}\label{p2}
\begin{align}
&\max \limits_{\widetilde{\boldsymbol{\psi}}}
~~\frac{2\Re\{\bar{a}a^H\}}{\bar{b}}-\frac{|\bar{a}|^2(b+|a|^2)}{\bar{b}(\bar{b}+|\bar{a}|^2)}+\nonumber\\
&~~~~~~~~\frac{2\Re\{\bar{\textbf{c}}^H\textbf{c}\}}{\bar{d}}-\frac{|\bar{\textbf{c}}|^2(d+|\textbf{c}|^2)}
{\bar{d}(\bar{d}+|\bar{\textbf{c}}|^2)}-\frac{1+e}{1+\bar{e}}\\
&~~\text{s.t.} ~~~|\widetilde{\boldsymbol{\psi}}(m)|\leq {\psi}^{\text{max}}, ~\widetilde{\boldsymbol{\psi}}(m+1)=1, ~ (\ref{phi_P}),
\end{align}
\end{subequations}
where
$a=\sqrt{\beta l P}\widetilde{\boldsymbol{\psi}}^H\textbf{h}_{bb}$, $b=\sigma^2_r\|\widetilde{\boldsymbol{\psi}}^H\textbf{H}_{bb}\|^2+\sigma^2_b,$
$\textbf{c}=(\sqrt{\sigma^2_r}\widetilde{\boldsymbol{\psi}}^H\textbf{H}_{ee})^H,$
$d=(1-\beta) l P \|\sqrt{g_{ae}}\textbf{h}^H_{ae}\textbf{T}_{AN}\|^2+\sigma^2_e,$
$e=\frac{\beta l P|\widetilde{\boldsymbol{\psi}}^H\textbf{h}_{ee}|^2+
\sigma^2_r\|\widetilde{\boldsymbol{\psi}}^H\textbf{H}_{ee}\|^2}{d},$
$\bar{a}$, $\bar{b}$, $\bar{\textbf{c}}$, $\bar{d}$, and $\bar{e}$ mean the solutions obtained at the previous iteration.
Then, the optimization problem (\ref{p2}) degenerate towards the following problem
\begin{subequations}\label{psi1}
\begin{align}
&\min \limits_{\widetilde{\boldsymbol{\psi}}}
~~\widetilde{\boldsymbol{\psi}}^H\textbf{W}\widetilde{\boldsymbol{\psi}}-2\Re\{\widetilde{\boldsymbol{\psi}}^H\textbf{u}\},\\
&~~\text{s.t.} ~~~|\widetilde{\boldsymbol{\psi}}(m)|\leq {\psi}^{\text{max}}, ~\widetilde{\boldsymbol{\psi}}(m+1)=1, ~ (\ref{phi_P}),
\end{align}
\end{subequations}
where
\begin{align}
&\textbf{W}=\frac{|\bar{a}|^2}{\bar{b}(\bar{b}+|\bar{a}|^2)}
(\beta l P\textbf{h}_{bb}\textbf{h}^H_{bb}+\sigma^2_r\textbf{H}_{bb}\textbf{H}^H_{bb})+\frac{|\bar{\textbf{c}}|^2}{\bar{d}(\bar{d}+|\bar{\textbf{c}}|^2)}\cdot\nonumber\\
&~~~~~~\sigma^2_r\textbf{H}_{ee}\textbf{H}^H_{ee}+
\frac{1}{1+\bar{e}}\frac{\beta l P\textbf{h}_{ee}\textbf{h}^H_{ee}+
\sigma^2_r\textbf{H}_{ee}\textbf{H}^H_{ee}}{{d}},\\
&\textbf{u}=\frac{1}{\bar{b}}\beta l P\textbf{h}_{bb}\textbf{h}^H_{bb}\widetilde{\boldsymbol{\psi}}_t+
\frac{1}{\bar{d}}\sigma^2_r\textbf{H}_{ee}\textbf{H}^H_{ee}\widetilde{\boldsymbol{\psi}}_t,
\end{align}
and $\widetilde{\boldsymbol{\psi}}_t$ stands for the solution obtained at the previous iteration.
It is noted that the problem (\ref{psi1}) is convex, which can be derived directly with CVX.

\begin{figure*}[htbp]
 \setlength{\abovecaptionskip}{-5pt}
 \setlength{\belowcaptionskip}{-10pt}
 \centering
 \begin{minipage}[t]{0.33\linewidth}
  \centering
\includegraphics[width=2.56 in]{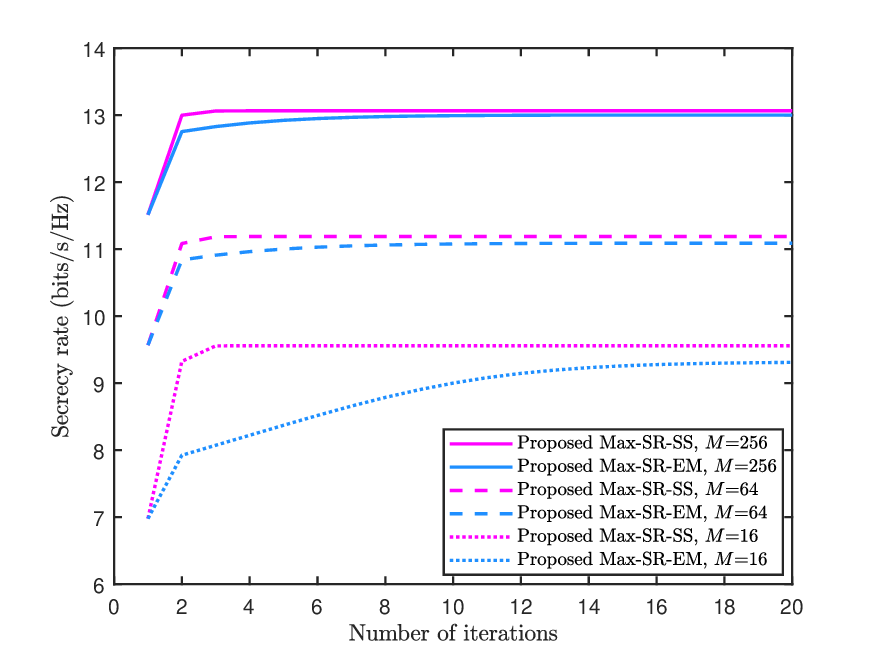}\\
\caption{Convergence of proposed schemes.}\label{multiple_itea}
 \end{minipage}%
 \begin{minipage}[t]{0.33\linewidth}
  \centering
\includegraphics[width=2.56 in]{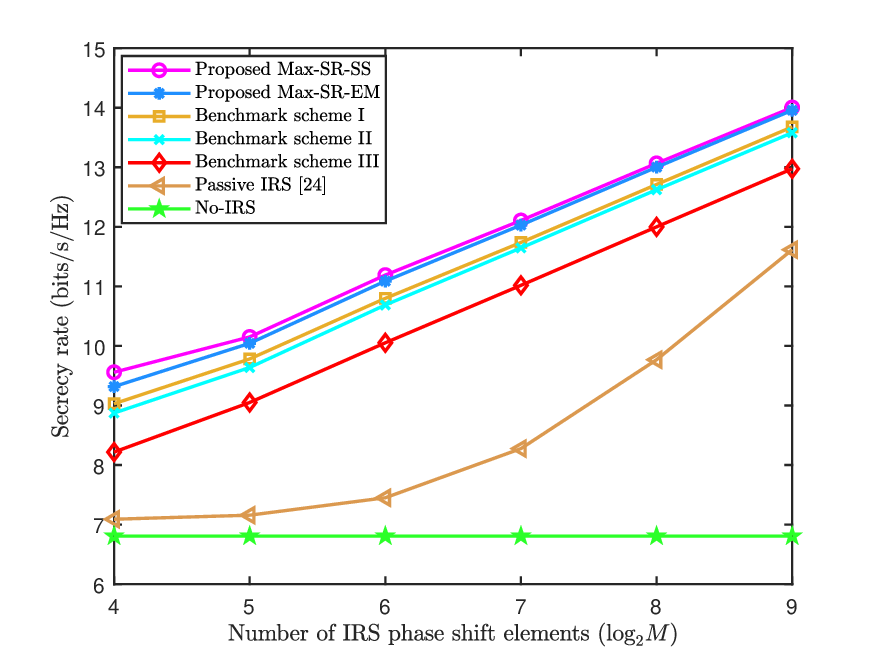}\\
\caption{SR versus the number of IRS elements $M$.}\label{multiple_M}
 \end{minipage}
 \begin{minipage}[t]{0.33\linewidth}
  \centering
\includegraphics[width=2.56 in]{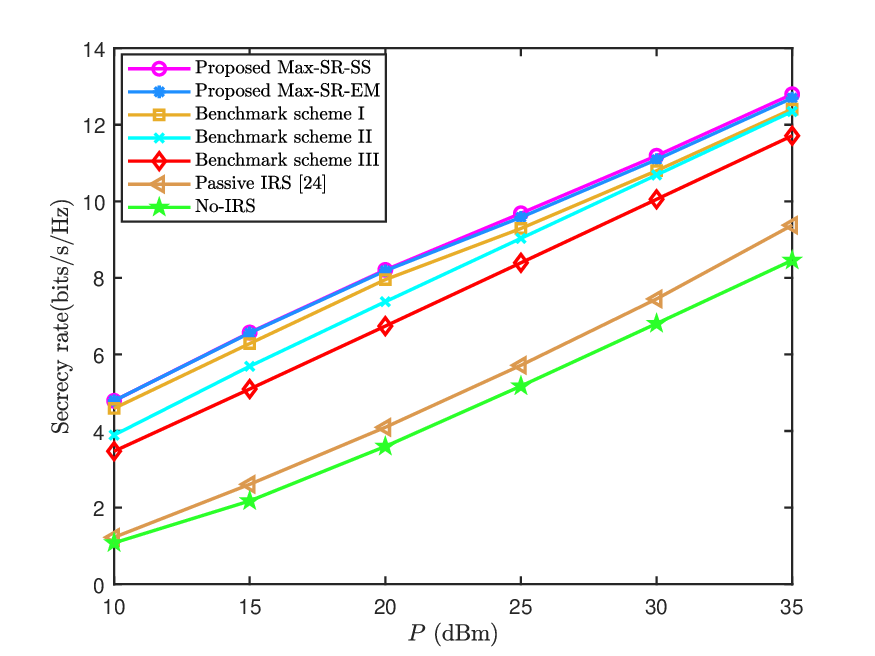}\\
\caption{SR versus the total power $P$.}\label{Multiple_P}
 \end{minipage}
\end{figure*}

\subsection{Overall scheme and complexity analysis}\label{}
So far, we have completed the derivation of the PA factors $\beta$ and $l$, beamforming vector $\textbf{v}$, and IRS phase shift matrix $\boldsymbol{\Psi}$. To make the procedure of this scheme clearer, we summarize the whole proposed {Max-SR-AO} algorithm below. (1) Initialize $\beta$, $l$, $\textbf{v}$, and $\boldsymbol{\Psi}^{(0)}$ to feasible solutions; (2) fixing  $l$, $\textbf{v}$, and $\boldsymbol{\Psi}$, solve (\ref{A1}) to update $\beta$; (3) given $\beta$, $\textbf{v}$, and $\boldsymbol{\Psi}$, solve (\ref{L1}) to update $l$; (4) fix $\beta$, $l$, and $\boldsymbol{\Psi}$, optimize (\ref{v3}) to update $\textbf{v}$; (5) given $\beta$, $l$, and $\textbf{v}$, solve (\ref{psi1}) to update $\widetilde{\boldsymbol{\psi}}$, and $\boldsymbol{\Psi}=\text{diag}\{\widetilde{\boldsymbol{\psi}}(1:M)\}^*$. Optimize the four variables alternately until the termination condition is satisfied.


In this scheme, the objective value sequence $\{R_s(\beta^{(k)}, l^{(k)}, \textbf{v}^{(k)}, \boldsymbol{\Psi}^{(k)})\}$ obtained in each iteration of the alternate optimization strategy is non-decreasing, and $R_s(\beta^{(k)}, l^{(k)}, \textbf{v}^{(k)}, \boldsymbol{\Psi}^{(k)})$ has a finite upper bound since the limited power constraint. Therefore, the convergence of the proposed Max-SR-AO scheme can be guaranteed.

The computational complexity of the overall {Max-SR-AO} algorithm is $\mathcal {O}\{L_{AO}[M^2\text{In}(1/\xi)+L_v(N^3+NM^2)+L_{\Psi}(2\sqrt{2}(M+1)^3+N(M+1)^2)]\}$
FLOPs, where $L_{AO}$ means the maximum number of alternating iterations, $L_v$ and $L_{\Psi}$ mean the iterative numbers of the subproblems (\ref{v3}) and (\ref{psi1}), respectively.

\section{Simulation Results}\label{s4}

To verify the performance of the proposed three maximum SR schemes, we perform the simulation comparison in this section. Unless otherwise noted, the parameters of the simulation are listed as follows: $P=30$dBm, $N=8$, $M=64$, $N_b=N_e=4$, $d_{ai}=110$m, $d_{ab}=d_{ae}=120$m, $\theta_{ai}=11\pi/36$, $\theta_{ab}=\pi/3$, $\theta_{ae}=5\pi/12$, $\sigma^2_b=\sigma^2_e=\sigma^2_r=-40$dBm. The path loss model is modeled as $g=\lambda^2/(4\pi d_{tr})^2$\cite{Emil2020Power}, where $\lambda$ and $d_{tr}$ stand for the wavelength and reference distance, respectively. For the sake of convenience, we set $(\lambda/(4\pi))^2=10^{-2}$. The convergence accuracy of the iterative scheme is set to be $\epsilon=10^{-3}$.

To evaluate the performance of the proposed schemes, the passive IRS scheme (i.e., GAI algorithm) in \cite{ShuEnhanced2021}, passive IRS scheme in \cite{Hu2020Directional}, passive IRS scheme (i.e., Algorithm 1) in \cite{Lin2023Enhanced}, and several benchmark schemes are applied for comparison at the same power, and these benchmark schemes are listed as follows.

1) \textbf{Benchmark scheme I:} Set the PA factor $l=0.6$, we only optimize the remaining variables alternatively.

2) \textbf{Benchmark scheme II:} Fixing the PA factor $\beta=0.5$, we only have to alternately optimize the rest variables.

3) \textbf{Benchmark scheme III:} Both the PA factors $\beta$ and $l$ are fixed at 0.5, i.e., $\beta=l=0.5$, and only the residual variables need to be optimized alternately.

4) \textbf{No-IRS:} Set all the active IRS related channel vectors and matrix to zero vectors and zero matrix, then, we only have to optimize the remaining variables alternatively.

\subsection{Bob and Eve are equipped with multiple antennas}
\begin{figure*}
 \setlength{\abovecaptionskip}{-5pt}
 \setlength{\belowcaptionskip}{-10pt}
 \centering
 \begin{minipage}[t]{0.33\linewidth}
  \centering
\includegraphics[width=2.56 in ]{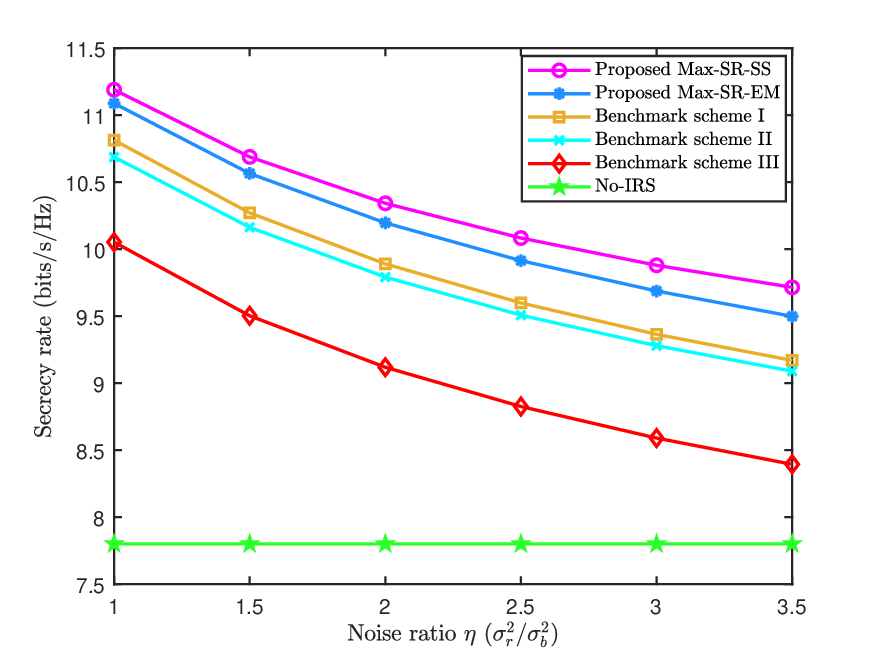}\\
\caption{SR versus the noise ratio $\eta$.}\label{multiple_sigma}  
 \end{minipage}%
 \begin{minipage}[t]{0.33\linewidth}
  \centering
\includegraphics[width=2.56 in]{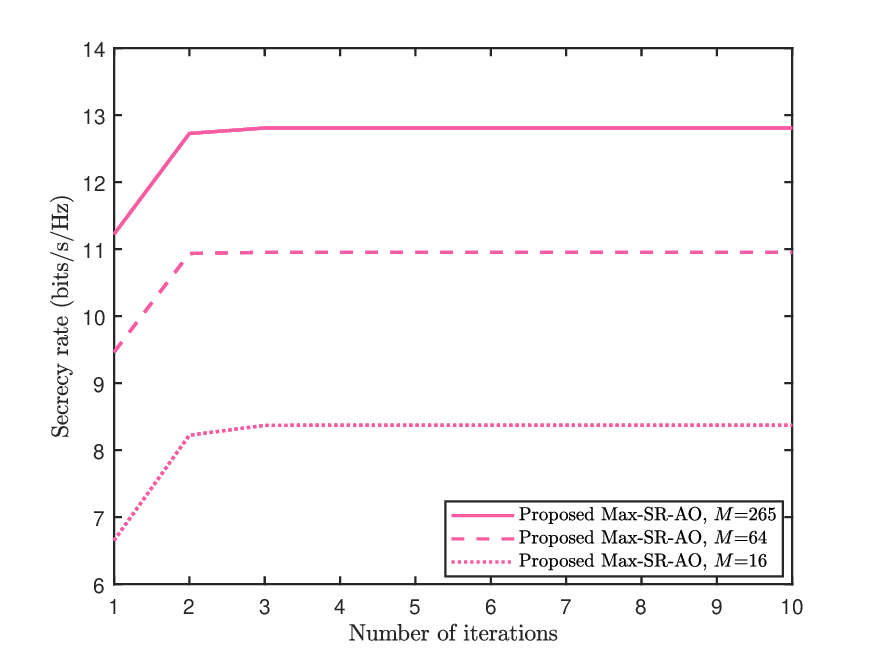}\\
\caption{Convergence of proposed scheme.}\label{single_itea}
 \end{minipage}
 \begin{minipage}[t]{0.33\linewidth}
  \centering
\includegraphics[width=2.56 in]{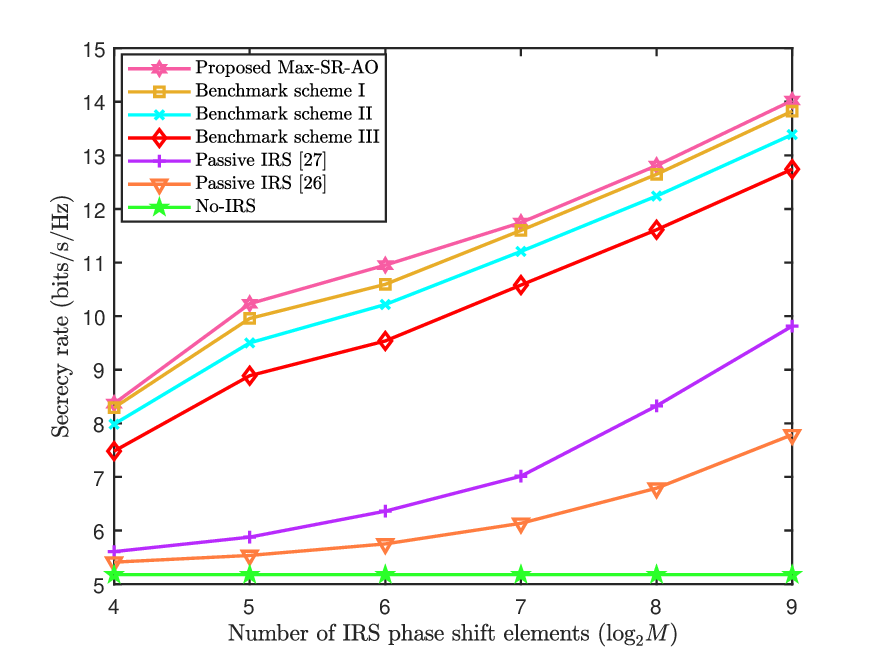}\\
\caption{SR versus the number of IRS elements $M$.}\label{single_M}
 \end{minipage}
\end{figure*}

Firstly, we show the convergence of both the proposed alternating optimization schemes in Fig.~\ref{multiple_itea}, where the number of phase shift elements of IRS $M=16, 64, 256$. It can be seen from the figure that the SRs of both proposed schemes increase rapidly with the number of iterations and finally converge to a value after a finite number of iterations. And the convergence speed of the proposed Max-SR-SS scheme is slightly faster than that of the proposed Max-SR-EM scheme. In addition, the SRs of both proposed schemes increase with the increases of $M$, and the SR of the proposed Max-SR-SS scheme is slightly better than that of the proposed Max-SR-EM scheme, regardless of the values of $M$. Combined with the previous analysis of the computational complexity of both, it can be found that the low-complexity of the latter is achieved at the price of some performance loss. As a result, the proposed Max-SR-EM scheme strikes a good balance between computational complexity and SR performance.




Fig.~\ref{multiple_M} plots the curves of the SR versus the number $M$ of active IRS phase shift elements of the proposed two schemes and benchmark schemes. Observing this figure, it can be found that the SRs of both the proposed schemes and benchmark schemes gradually increase with the increases of $M$, they have a decreasing order in terms of SR performance: proposed Max-SR-SS, proposed Max-SR-EM, benchmark scheme I, benchmark scheme II, benchmark scheme III, passive IRS \cite{ShuEnhanced2021}, and no IRS. The SR difference between the two proposed schemes is trivial with the increases of $M$, and they make significant SR performance enhancements over the five benchmark schemes at the same total power budget. For example, when $M=64$, the SR performance enhancements achieved by both the proposed schemes over the benchmark scheme I, benchmark scheme II, benchmark scheme III, passive IRS \cite{ShuEnhanced2021}, and no IRS are above $3\%$, $4\%$, $11\%$, $40\%$, and $47\%$, respectively. These further explain the motivation for investigating the active IRS, PA, and beamforming algorithms.



Fig.~\ref{Multiple_P} depicts the curves of the SR versus the total power $P$ ranging from 10dBm to 35dBm. From this figure, we can learn that the SRs of two proposed schemes and five benchmark schemes increase with the increases of $P$, and the ordering of their achieved SRs is similar to that of Fig.~\ref{multiple_M}. The difference in SR performance between proposed Max-SR-SS scheme and benchmark scheme I is slightly less than that between it and benchmark scheme II, which means that optimizing the confidential message PA factor $\beta$ has a more significant performance enhancement for the system compared to optimizing the base station PA factor $l$ in this paper. Compared to the benchmark schemes of no IRS and passive IRS \cite{ShuEnhanced2021}, the SRs achieved by the both proposed schemes and the remaining benchmark schemes are remarkable, with the latter being more than one times higher than the former. This is because active IRS elements equipped with power amplifiers enable more SR performance gain. Moreover, the gap between the SRs of the two proposed schemes is trivial when $P\leq20$dBm.


Fig.~\ref{multiple_sigma} demonstrates the curves of the SR versus the noise ratio $\eta$ ranging from 1 to 3.5, where $\eta=\sigma^2_r/\sigma^2_b$ and $\sigma^2_b$ remains constant, i.e., the increase of $\eta$ is equivalent to that of the noise power at the active IRS. This figure shows that apart from the scheme of no IRS, the SRs of two proposed schemes and the benchmark schemes I $\sim$ III decrease gradually with the increases of $\eta$. This is due to the fact that the active IRS helps to transmit the confidential information to Bob and also reflects the noise generated at the IRS to him. When $\eta$ increases, the noise received by Bob also increases, which leads to a decrease in the SR performance for all schemes apart from the no IRS scheme. Taking Max-SR-SS scheme as an example, the SR at $\eta=2$ and $\eta=3$ are above 8\% and 13\% lower than those at $\eta=1$, respectively.


\subsection{Bob and Eve are equipped with single antenna}

Fig.~\ref{single_itea} shows the SR versus the number of iterations of the proposed Max-SR-AO scheme. It can be seen from this figure that regardless of the value of $M$, the proposed Max-SR-AO scheme takes about four iterations to converge the SR ceiling. Fig.~\ref{single_M} plots the SR versus the number $M$ of the IRS phase shift elements. It can be found that similar to the scenario where both Bob and Eve are equipped with multiple antennas, the SR performance of the proposed Max-SR-AO scheme is slightly better than that of the fixed PA schemes and significantly better than that of the passive IRS \cite{Lin2023Enhanced}, passive IRS \cite{Hu2020Directional}, and no IRS schemes.

\begin{figure}[htbp]
\centering
\includegraphics[width=0.35\textwidth]{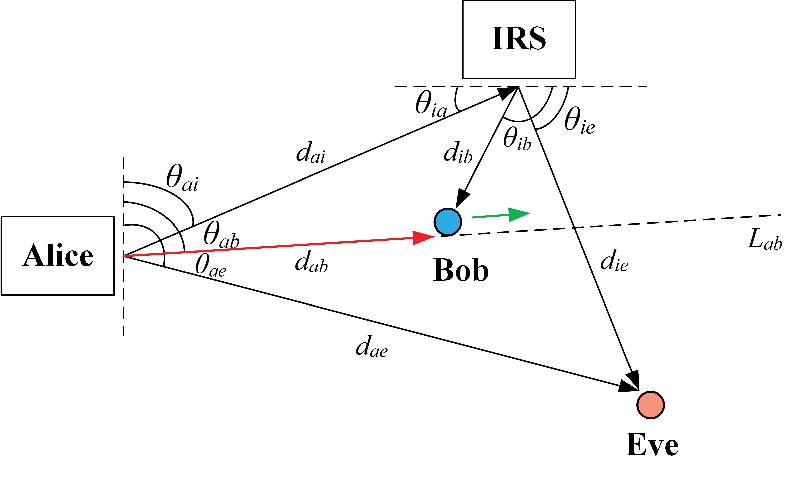}\\
\caption{Diagram of Bob's movement.}\label{move}
\end{figure}

To investigate the impact of the Bob's location on SR performance, with fixed positions of Alice, IRS, and Eve, we assume that Bob moves only along the straight line $L_{ab}$ (i.e., the line connecting Alice and Bob) for simplicity of analysis. At this point, the Bob's location only depends on the distance $d_{ab}$ of Alice-to-Bob link. As $d_{ab}$ gradually increases, Bob first moves closer to the IRS, reaches a peak and then moves away from it. The diagram of Bob's detailed movement as shown in Fig.~\ref{move}.

\begin{figure}[htbp]
\centering
\includegraphics[width=0.35\textwidth]{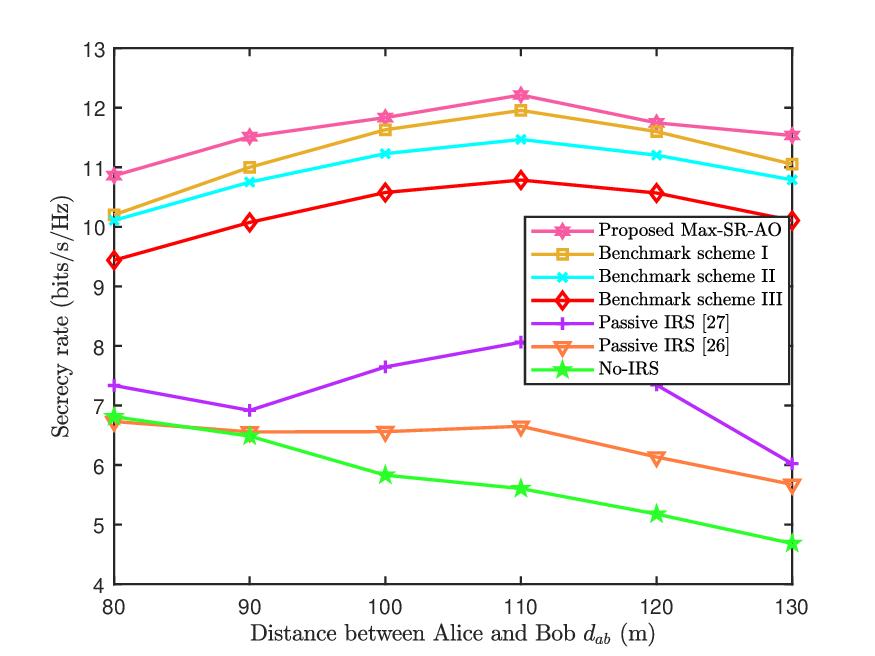}\\
\caption{SR versus the distance between Alice and Bob $d_{ab}$.}\label{single_distance}
\end{figure}

Based on the model of Bob's position movement in Fig.~\ref{move}, Fig.~\ref{single_distance} presents the curves of the SR versus the distance $d_{ab}$ ranging from 80m to 130m, where $M=128$. It reveals that as Bob's position moves away from Alice along $L_{ab}$ and closer to the IRS, the SR of the no-IRS scheme gradually decreases with the increase of $d_{ab}$. For the proposed Max-SR-AO scheme, first, when Bob is positioned between Alice and IRS and away from them, its energy received from Alice gradually decreases and its SRs gradually decreases with increasing $d_{ab}$. Then, as Bob moves away from Alice and closer to the IRS, their energy received from the IRS gradually increases and their SRs gradually increase and reach a peak when Bob is at the bottom of the IRS. Finally, with Bob moving away from Alice and IRS, their energy from Alice and IRS gradually decreases and the SRs gradually decrease. Moreover, there are similar SR performance tendencies for passive IRS \cite{Hu2020Directional}, and passive IRS \cite{Lin2023Enhanced}. After the peak, the gap of SRs gained by the proposed schemes and passive IRS schemes increases gradually with $d_{ab}$. Furthermore, the proposed scheme has better SRs performance than the benchmark schemes I $\sim$ III regardless of the value of $d_{ab}$, which highlights the significance of optimizing the PA factors.

\section{Conclusion}\label{s5}
In this paper, we made an investigation of active IRS-aided DM network and focused on adjusting the PA between IRS and Alice to improve the SR performance. To the best of our knowledge, such a PA has not been investigated the optimization of the PA factors, transmit and receive beamforming, and phase shift matrix of IRS in the active IRS-assisted DM network. Firstly, to maximize SR with AN only interfering with Eve, the projection matrix of AN was designed based on the criterion of null-space projection. Then, to address the formulated maximum SR optimization problem, two alternating iteration schemes, namely Max-SR-SS and Max-SR-EM, were proposed. The former with a high-performance employed the derivative operation, SCA, and generalized Rayleigh-Rize methods to find the optimal PA factors, transmit and receive beamforming, and IRS phase shift matrix. While the latter with a low-complexity got the closed-form expression of the IRS phase shift matrix by the criteria of EAR and MM. Moreover, a special case of receivers equipped with single antenna was considered, and a Max-SR-AO scheme was proposed to address the problem. Simulation results showed that the SR of the DM network was dramatically enhanced with the help of active IRS compared to the passive IRS scheme, and the proposed joint PA and beamforming schemes have made an obvious SR enhancement over the schemes with fixed PA.



\section*{Appendix}

Let us define
$\textbf{q}=\sqrt{\beta l P}(\sqrt{g_{ae}}\textbf{H}^H_{ae}+\sqrt{g_{aie}}\textbf{H}^H_{ie}\boldsymbol{\Psi}\textbf{H}_{ai})\textbf{v},$
$\textbf{Q}_1=\textbf{q}\textbf{q}^H,$
$\textbf{Q}_2=(1-\beta) l Pg_{ae}\textbf{H}^H_{ae}\textbf{T}_{AN}\textbf{T}^H_{AN}\textbf{H}_{ae}+
\sigma^2_rg_{ie}\textbf{H}^H_{ie}\boldsymbol{\Psi}\boldsymbol{\Psi}^H\textbf{H}_{ie}+\sigma^2_e\textbf{I}_{N_e},$
$\textbf{Q}_2=\textbf{Q}_2^{1/2}(\textbf{Q}_2^{1/2})^H,$
$\textbf{w}=\textbf{Q}_2^{1/2}\textbf{u}_e,$
then, $\textbf{u}_e=\textbf{Q}_2^{-1/2}\textbf{w}$, and
\begin{align}\label{Q}
\gamma=\frac{\textbf{u}_e^H\textbf{Q}_1\textbf{u}_e}{\textbf{u}_e^H\textbf{Q}_2\textbf{u}_e}=
\frac{\textbf{w}^H(\textbf{Q}_2^{-1/2})^H\textbf{Q}_1\textbf{Q}_2^{-1/2}\textbf{w}}{\textbf{w}^H\textbf{w}}.
\end{align}
$\widetilde{\textbf{w}}=\frac{\textbf{w}}{\|\textbf{w}\|},$
we have $\textbf{w}=\widetilde{\textbf{w}}\|\textbf{w}\|$.
Then, (\ref{Q}) can be rewritten as
\begin{align}
&\frac{\textbf{w}^H(\textbf{Q}_2^{-1/2})^H\textbf{Q}_1\textbf{Q}_2^{-1/2}\textbf{w}}{\textbf{w}^H\textbf{w}}\nonumber\\
&=\frac{\widetilde{\textbf{w}}^H\|\textbf{w}\|(\textbf{Q}_2^{-1/2})^H\textbf{Q}_1\textbf{Q}_2^{-1/2}\|\textbf{w}\|\widetilde{\textbf{w}}}
{\|\textbf{w}\|^2\widetilde{\textbf{w}}^H\widetilde{\textbf{w}}}\nonumber\\
&=\widetilde{\textbf{w}}^H(\textbf{Q}_2^{-1/2})^H\textbf{Q}_1\textbf{Q}_2^{-1/2}\widetilde{\textbf{w}}\nonumber\\
&\leq \lambda_{\text{max}}((\textbf{Q}_2^{-1/2})^H\textbf{Q}_1\textbf{Q}_2^{-1/2})\nonumber\\
&\buildrel (a) \over=\text{Tr}(\textbf{Q}_2^{-1}\textbf{Q}_1),
\end{align}
(a) is due to the fact that
$\text{rank}((\textbf{Q}_2^{-1/2})^H\textbf{Q}_1\textbf{Q}_2^{-1/2})
=\text{rank}((\textbf{Q}_2^{-1/2})^H\textbf{q}\textbf{q}^H\textbf{Q}_2^{-1/2})
=1.
$

\ifCLASSOPTIONcaptionsoff
  \newpage
\fi

\bibliographystyle{IEEEtran}

\bibliography{IEEEfull,reference}
\end{document}